\documentclass{article}
\pdfoutput=1
\usepackage[top=2cm,bottom=3cm,left=2cm,right=2cm]{geometry}
\usepackage{amsmath}
\usepackage{amsfonts}
\usepackage{stmaryrd}
\usepackage{bm}
\usepackage{hyperref}
\usepackage{upgreek}
\usepackage[mathscr]{euscript}
\usepackage{graphicx}
\usepackage{mathtools}
\usepackage{soul}

\numberwithin{equation}{section}
\numberwithin{figure}{section}

\def\eq#1{(\ref{eq:#1})}
\def\lineup{\!\!\!\!\!\!\!\!\!\!&&}

\def\d{\partial}

\def\H{\mathcal{H}}

\def\[{\big[}
\def\]{\big]}

\begin{document}

\begin{titlepage}
\rightline\today

\begin{center}
\vskip 3.5cm

{\large \bf{Vertical Integration from the Large Hilbert Space}}

\vskip 1.0cm

{\large {Theodore Erler\footnote{tchovi@gmail.com}, Sebastian Konopka\footnote{konopka.seb@googlemail.com}}}

\vskip 1.0cm

{\it Institute of Physics of the ASCR, v.v.i.}\\
{\it Na Slovance 2, 182 21 Prague 8, Czech Republic}\\




\vskip 2.0cm

{\bf Abstract} 

\end{center}

We develop an alternative description of the procedure of vertical integration based on the observation that amplitudes can be written in BRST exact form in the large Hilbert space. We relate this approach to the description of vertical integration given by Sen and Witten. 

\end{titlepage}

\tableofcontents

\section{Introduction and Summary}

Computing a superstring scattering amplitude requires inserting a configuration of picture changing operators (PCOs) for each Riemann surface contributing to the amplitude. A choice of PCOs roughly corresponds to a section of a fiber bundle: The base of the fibre bundle consists of the moduli space of Riemann surfaces with the relevant genus, spin structure, and number of punctures, and the fiber at each point consists of $p$ copies of the Riemann surface with the corresponding value of the moduli; this parameterizes the possible ways of inserting $p$ PCOs on that surface.  The worldsheet path integral defines a measure on this fiber bundle which can be pulled back to any submanifold; in particular, pulling the measure back onto a section of the fiber bundle and integrating defines a superstring amplitude with a prescribed configuration of PCOs on each Riemann surface. A significant complication with this procedure, however, is the appearance of spurious singularities in the superstring measure \cite{Verlinde}. One can try to look for a global section which avoids spurious singularities everywhere, but this may be inconvenient in practice, or there may be an obstruction to the existence of such a global section.\footnote{It is known that the supermoduli space of super-Riemann surfaces cannot be holomorphically projected down to the ordinary (bosonic) moduli space of Riemann surfaces \cite{Donagi}. To our knowledge, the implications of this fact  from the point of view of PCOs has not been worked out, but one possibility is that the PCO positions cannot be chosen globally as holomorphic functions of the moduli. } 

A remedy proposed by Sen \cite{SenOffShell}, and later made more explicit by Sen and Witten \cite{SenWitten}, is to divide the moduli space into regions so that on each region we can choose a local section which avoids spurious poles. Simply adding the contributions from the local sections together, however, does not define a gauge invariant amplitude. To correct for this, at the interface between the different regions of the moduli space we must integrate ``along the fiber" to connect local sections---that is, one must deform one choice of PCOs into another while keeping the moduli fixed. This is called {\it vertical integration}. The amplitude is then defined by a closed integration cycle in the fiber bundle composed of local sections connected by ``vertical segments." Importantly, the nature of the superstring measure implies that spurious singularities can be rendered harmless on the vertical segments. Therefore we can obtain gauge invariant amplitudes free from unphysical divergences in the measure.

In this paper we investigate a different, more algebraic, approach to this problem, motivated by recent studies of superstring field theories \cite{WittenSS,ClosedSS}. Consider an $n$-point amplitude expressed as an $n$-fold bra state:
\begin{equation}\langle A^{p}|:\ \mathcal{H}^{\otimes n}\to \mathbb{C},\end{equation}
where $\mathcal{H}$ is a CFT vector space containing BRST invariant physical states. The superscript indicates that the amplitude contains $p$ picture changing operators inserted in some way on the constituent Riemann surfaces. Gauge invariance of the amplitude is equivalent to the statement that this bra state is annihilated by a sum of BRST operators acting on each external state: 
\begin{equation}\langle A^{p}|\Big(Q\otimes\mathbb{I}^{\otimes n-1} +...+\mathbb{I}^{\otimes n-1}\otimes Q\Big) = 0.\end{equation}
Here $Q$ denotes the BRST operator and $\mathbb{I}$ is the identity operator on $\mathcal{H}$. It is well-known that the cohomology of $Q$ is trivial in the ``large Hilbert space" introduced by Friedan, Martinec, and Shenker \cite{FMS}, that is, the CFT state space obtained by bosonizing the $\beta\gamma$ ghosts into the $\eta,\xi,e^\phi$ system and allowing for states which depend on the zero mode of the $\xi$ ghost. This implies that the amplitude can be expressed in the form 
\begin{equation}\langle A^{p}| = \langle \alpha^{p}|\Big(Q\otimes\mathbb{I}^{\otimes n-1} +...+\mathbb{I}^{\otimes n-1}\otimes Q\Big).\label{eq:ApQ}\end{equation}
We will call the $n$-fold bra state $\langle \alpha^{p}|$ a {\it gauge amplitude}, following the terminology of \cite{ClosedSS}. The gauge amplitude lives in the large Hilbert space. However, the physical amplitude $\langle A^{p}|$ must reside in the ``small Hilbert space" where the zero mode of the $\xi$ ghost is absent. This requires that the amplitude satisfies 
\begin{equation}\langle A^{p}|\Big(\eta\otimes\mathbb{I}^{\otimes n-1} +...+\mathbb{I}^{\otimes n-1}\otimes \eta\Big) = 0,
\end{equation}
where $\eta$ denotes the zero mode of the eta ghost. This is consistent with \eq{ApQ} provided that the object
\begin{equation}\langle\alpha^{p}|\Big(\eta\otimes\mathbb{I}^{\otimes n-1} +...+\mathbb{I}^{\otimes n-1}\otimes \eta\Big)\end{equation}
is annihilated by the BRST operator. Since $\eta$ carries picture $-1$, it is natural to interpret this object as an amplitude containing one fewer PCO insertion:
\begin{equation}\langle\alpha^{p}|\Big(\eta\otimes\mathbb{I}^{\otimes n-1} +...+\mathbb{I}^{\otimes n-1}\otimes \eta\Big) = \langle A^{p-1}|.\end{equation}
We can now apply this procedure again, expressing $\langle A^{p-1}|$ as the BRST variation of a gauge amplitude $\langle \alpha^{p-1}|$, and apply $\eta$ once again to arrive at an amplitude $\langle A^{p-2}|$ containing two fewer PCO insertions. Continuing this process unfolds a hierarchy of amplitudes and gauge amplitudes:
\begin{eqnarray}
\langle A^{p}|\lineup = \langle \alpha^{p}|\Big(Q\otimes\mathbb{I}^{\otimes n-1} +...+\mathbb{I}^{\otimes n-1}\otimes Q\Big)\nonumber\\
\langle A^{p-1}| \lineup = \langle \alpha^{p}|\Big(\eta\otimes\mathbb{I}^{\otimes n-1} +...+\mathbb{I}^{\otimes n-1}\otimes \eta\Big)\nonumber\\
\langle A^{p-1}|\lineup = \langle \alpha^{p-1}|\Big(Q\otimes\mathbb{I}^{\otimes n-1} +...+\mathbb{I}^{\otimes n-1}\otimes Q\Big)\nonumber\\
\lineup\vdots\phantom{\Big(}\nonumber\\
\langle A^{1}| \lineup = \langle \alpha^{2}|\Big(\eta\otimes\mathbb{I}^{\otimes n-1} +...+\mathbb{I}^{\otimes n-1}\otimes \eta\Big)\nonumber\\
\langle A^{1}|\lineup = \langle \alpha^{1}|\Big(Q\otimes\mathbb{I}^{\otimes n-1} +...+\mathbb{I}^{\otimes n-1}\otimes Q\Big)\nonumber\\
 \langle A^{0}| \lineup = \langle \alpha^{1}|\Big(\eta\otimes\mathbb{I}^{\otimes n-1} +...+\mathbb{I}^{\otimes n-1}\otimes \eta\Big),
\end{eqnarray}
where at the end we obtain an amplitude $\langle A^{0}|$ containing no PCO insertions at all.\footnote{The number of PCO insertions in $\langle A^{p}|$ is determined by the requirement that the amplitude is nonzero acting on NS states at picture $-1$ and Ramond states at picture $-1/2$. This means that the  amplitudes with fewer than $p$ PCO insertions will need to act on states with nonstandard picture to obtain a nonzero result. Generally, such amplitudes will encounter divergences from spurious singularities. We discuss such amplitudes formally, as intermediate objects used to obtain the final amplitude $\langle A^{p}|$ which is gauge invariant and free from spurious singularity.} The structure here is reminiscent of descent equations which appear in analysis of anomalies in gauge theories \cite{descent}. This leads the following procedure for deriving gauge invariant amplitudes. First we start with the amplitude $\langle A^{0}|$, and insert the operator $\xi(z)$ at some point on each constituent Riemann surface. This defines the gauge amplitude $\langle \alpha^{1}|$. We then take the BRST variation to arrive at the amplitude $\langle A^{1}|$ containing one PCO. We then insert another $\xi(z)$ on the Riemann surfaces of $\langle A^{1}|$ to derive $\langle \alpha^{2}|$, and continue in this way until we arrive at the amplitude $\langle A^{p}|$ containing $p$ PCO insertions. The crucial point is that the insertions of $\xi$ do not need to vary continuously with the moduli to ensure gauge invariance. Gauge invariance is automatic since the final amplitude $\langle A^{p}|$ is expressed in BRST exact form. We may therefore allow the $\xi$ insertions to ``jump" across spurious poles discontinuously as a function of the moduli to avoid unphysical divergences. 

The primary goal of this work is to show that the above algebraic procedure gives a viable alternative to defining a consistent measure on the moduli space for superstring amplitudes. The approach has some advantages. The computation of vertical corrections is arguably simpler and more flexible, and certain essential properties, such as gauge invariance of the amplitude and independence from various choices, are evident from the nature of the construction. A second goal of our work will be understanding the relationship between the algebraic approach and the conceptually rather different idea of vertical integration. Our motivation is to form a link between Sen's discussion of superstring field theories~\cite{SenRev} and other techniques which have been independently developed based on the large Hilbert space \cite{WittenSS,ClosedSS,complete}. As investigations continue into quantum effects in superstring field theories \cite{hyperbolic,tadpole}, and in the geometrical formulation based on super-Riemann surfaces~\cite{Ohmori}, it may be useful to have an understanding of the relationship between these approaches. 

Vertical integration is a general idea which can be implemented in many ways. To give ourselves a concrete objective, we will focus on the connection to the vertical integration procedure as implemented by Sen and Witten~\cite{SenWitten}. In the spirit of that work, we discuss only on-shell amplitudes and ignore the fact that the moduli space of Riemann surfaces is noncompact. The boundary of moduli space is associated with the infrared physics of superstring perturbation theory, about which there has been extensive discussion in recent years. One way of dealing with infrared divergences is to extend amplitudes off-shell using the formalism of string field theory. Suffice it to say that our discussion can be easily adapted in this context, which provided part of the motivation for this work. For simplicity we discuss PCOs in the holomorphic sector only, as would be relevant for the heterotic string. For type II strings we have a similar story also in the antiholomorphic sector.

This paper is organized as follows. In section \ref{sec:measure} we review the definition of superstring measure in the PCO formalism. In section \ref{sec:alg} we describe the algebraic construction of superstring amplitudes, deriving a set of recursive equations for the vertical corrections at the interface between local sections needed to ensure gauge invariance. We give examples and prove that on-shell amplitudes are independent of the choice of vertical corrections derived by this procedure. In section \ref{sec:SenWitten} we discuss the construction of Sen and Witten. To give a clear formalization of their procedure, we employ an analogue of differential forms on the lattice, called {\it difference forms}. The Sen-Witten vertical corrections are defined by ``integration" of a discretized measure---characterized by difference forms---over a collection of links in a $p$-dimensional cubical lattice, where $p$ is the number of PCOs in the amplitude. The sites of the lattice correspond to combinations of PCOs taken from adjoining local sections, and the collection of links which define the ``integration cycle" are called {\it lattice chains}. We describe how vertical corrections of this form may be constructed from the algebraic point of view. The algebraic construction introduces a collection of auxiliary amplitudes containing $0,...,p-1$ PCOs whose vertical corrections are characterized by lattice chains inside lower dimensional lattices of respective dimension $0,...,p-1$. The algebraic construction functions by extending lattice chains from lower dimensional into higher dimensional lattices in such a way as to be consistent with gauge invariance and so that the chains of higher dimensional lattices project down to the chains of lower dimensional lattices. We conclude with some examples.

\section{Superstring Measure}
\label{sec:measure}

In this section we review the superstring measure in the PCO formalism \cite{SenOffShell}. The purpose is to fix a convenient notation for our calculations and to simplify some signs. 

Given a Riemann surface with genus $g$ and $n$ punctures, we can remove $n$ disks around each puncture and cut what remains into $2g+n-2$ components with the topology of a sphere with three holes. We cover the $n$ disks with holomorphic local coordinates $w_1,...,w_n$ with $|w_a|<1$. The origin of these coordinates $w_a=0$ corresponds to the location of the punctures on the Riemann surface. On the $2g+n-2$ spheres we introduce holomorphic coordinates~$z_i$. The Riemann surface can be reconstructed by gluing the boundaries of these components with holomorphic transition functions:
\begin{eqnarray}
z_i\lineup =f_{ij}(z_j)\label{eq:ztrans}\\
z_i\lineup =f_a(w_a).\label{eq:wtrans}
\end{eqnarray}
The transition functions exist between coordinates which are identified by the gluing, and encode all information about the moduli of the Riemann surface. Since we work with the heterotic string, we are interested in the moduli space of Riemann surfaces with spin structure in the leftmoving sector. This is a $4^g$-fold covering of the bosonic moduli space which comes in two disconnected components, representing the even and odd spin structures. We use $M$ to denote one of these disconnected components. That is, $M$ is the moduli space of genus $g$ Riemann surfaces with $n$ punctures together with either an even or odd spin structure, and $m\in M$ denotes a point in this moduli~space. 

Given transition functions $f_a,f_{ij}$, we may define an $n$-fold bra state called a {\it surface state} 
\begin{equation}\langle \Sigma|:\H^{\otimes n}\to \mathbb{C}.\end{equation}
The surface state is defined so that the quantity
\begin{equation}
\langle \Sigma|\Phi_1\otimes...\otimes\Phi_n\label{eq:Sigcorr}
\end{equation}
represents a correlation function on a Riemann surface assembled with the transition functions $f_a,f_{ij}$, with the vertex operators corresponding to the states $\Phi_1,...,\Phi_n$ inserted at the punctures in the respective coordinates $w_1,...,w_n$. If the vertex operators are conformally invariant, the correlation function only depends on the transition functions through the moduli of the Riemann surface they represent. For generic vertex operators, the surface state will depend more nontrivially on the choice of transition functions. The surface state is BRST invariant:
\begin{equation}\langle \Sigma|Q = 0.\end{equation}
Since this will not cause confusion, we use a shorthand notation where $Q$ represents a sum of BRST operators acting on each state:
\begin{equation}Q\ \rightarrow \ Q\otimes\mathbb{I}^{\otimes n-1} +...+\mathbb{I}^{\otimes n-1}\otimes Q.\end{equation}
In particular $\langle \Sigma|Q$ represents a correlation function with a contour integral of the BRST current surrounding all punctures. If we deform the contour inside the surface and shrink to a point, this gives zero. Also important for our discussion is the fact that $\langle \Sigma|$ is well-defined in the small Hilbert space. This implies that it is annihilated by the zero mode of the eta ghost: 
\begin{equation}\langle \Sigma|\eta = 0,\end{equation}
where $\eta$ denotes a sum of eta zero modes acting on each state. 

Let us first describe the measure without PCOs, as would be relevant for computing amplitudes in bosonic string theory. We fix a choice of surface state $\langle \Sigma(m)|$ for each point $m$ in the moduli space, and write $\langle \Sigma(m)|$ simply as $\langle\Sigma|$, leaving the dependence on moduli implicit. To define differential forms that can be integrated over the moduli space, we need to insert the appropriate $b$-ghosts inside correlation functions. Using the idea of  the {\it Schiffer variation}, following \cite{Zwiebach}, we may express the $b$-ghost insertions as contour integrals surrounding the punctures. Around the $a$th puncture we have a $b$-ghost insertion of the form
\begin{equation}b(v_\mu^{(a)})=\oint\frac{dw_a}{2\pi i} v_\mu^{(a)}(m,w_a) b(w_a) + \oint \frac{d\bar{w}_a}{2\pi i}\bar{v}_\mu^{(a)}(m,\bar{w}_a)\bar{b}(\bar{w}_a),\end{equation}
where the contours are oriented counterclockwise respectively in the $w_a$ and $\bar{w}_a$ coordinates on the Riemann surface. The contour integrals are weighted by functions $v_\mu^{(a)}(m,w_a)$ called {\it Schiffer vector fields}. The lower index $\mu$ corresponds to coordinates $m^\mu$ on the moduli space, with $\mu=1,...,6g+2n-6$. We introduce the operator 
\begin{eqnarray}
T_\mu\lineup \equiv \left(\oint\frac{dw_1}{2\pi i} v_\mu^{(1)}(m,w_1)T(w_1)\right)\otimes\mathbb{I}^{\otimes n-1}+...+\mathbb{I}^{\otimes n-1}\otimes \left(\oint\frac{dw_n}{2\pi i} v_\mu^{(n)}(m,w_n)T(w_n)\right)\nonumber\\
\lineup\ \ \ +\left(\oint\frac{d\bar{w}_1}{2\pi i} \bar{v}_\mu^{(1)}(m,\bar{w}_1)\bar{T}(\bar{w}_1)\right)\otimes\mathbb{I}^{\otimes n-1}+...+\mathbb{I}^{\otimes n-1}\otimes\left( \oint\frac{d\bar{w}_n}{2\pi i} \bar{v}_\mu^{(n)}(m,\bar{w}_n)\bar{T}(\bar{w}_n)\right).\label{eq:Tmu}
\end{eqnarray}
The Schiffer vector fields are defined so that the following equation holds: 
\begin{equation}\frac{\d}{\d m^\mu}\langle\Sigma| = -\langle \Sigma|T_\mu.\label{eq:SigmaT}\end{equation}
Additional properties are
\begin{eqnarray}
[Q,b_\mu]\lineup = T_\mu \\
\ [T_\mu,b_\nu] \lineup = \frac{\d}{\d m^\mu}b_\nu - \frac{\d}{\d m^\nu}b_\mu,\label{eq:Tb}
\end{eqnarray}
where $b_\mu$ is defined as in \eq{Tmu} with the energy momentum tensor replaced by the $b$-ghost, and $[\cdot,\cdot]$ represents a graded commutator with respect to Grassmann parity. Let $dm^\mu$ be coordinate 1-forms on the moduli space, and introduce operator-valued 1-forms:
\begin{eqnarray}
T \lineup \equiv dm^\mu T_\mu,\\
b\lineup \equiv dm^\mu b_\mu.
\end{eqnarray}
To simplify signs, we assume that the coordinate 1-forms $dm^\mu$ are uniformly Grassmann odd objects, so they anticommute through each other and also though Grassmann odd worldsheet operators. In this convention, the operator $T$ is Grassmann odd, and $b$ is Grassmann even. The identities \eq{SigmaT}-\eq{Tb} imply
\begin{eqnarray}
d\langle \Sigma|\lineup =-\langle \Sigma|T\label{eq:bosid1}\\
\ [Q,b]\lineup = -T\\
db\lineup = \frac{1}{2}[T,b],\label{eq:bosid3}
\end{eqnarray}
where
\begin{equation}d= dm^\mu\frac{\d}{\d m^\mu}
\end{equation}
is the exterior derivative on the moduli space. The measure for scattering amplitudes can then be expressed  
\begin{equation} \langle \Omega| = \langle \Sigma|e^b.
\end{equation}
This is a differential form of inhomogeneous degree. In particular, the operator $e^b$ is defined by the series expansion 
\begin{equation}
e^b = \mathbb{I}^{\otimes n} + b + \frac{1}{2!}b^2+...+\frac{1}{(6g+2n-6)!}b^{6g+2n-6}.
\end{equation}
The series terminates since there are only $6g+2n-6$ independent 1-forms $dm^\mu$. The last term is a top degree form, and this is the part of the measure that should be integrated over the moduli space to obtain the amplitude. Using the identities \eq{bosid1}-\eq{bosid3}, it is straightforward to show that 
\begin{equation}
\langle \Omega| Q = -d\langle \Omega|.\label{eq:bos_measure}
\end{equation}
Assuming we can ignore contributions from the boundaries of moduli space, this implies that BRST trivial states decouple from scattering amplitudes.

The measure $\langle \Omega|$, however, can only compute superstring scattering amplitudes between states of nonstandard picture. Such amplitudes will typically suffer from unphysical divergences due to spurious singularities. Therefore, it is useful to generalize the measure to accommodate correlation functions containing additional operator insertions, in particular PCOs. One concrete way to do this is as follows.\footnote{In the description of \cite{SenOffShell}, PCOs are inserted in the coordinates $z_i$ representing the Riemann surface with the disks around the punctures removed. In this approach, the Schiffer vector fields must be chosen to vanish at the location of the PCOs in order to ensure that deformations of the moduli are independent from deformations of the PCO positions in the coordinates $z_i$. This is equivalent to the approach we take, but expressed in a different coordinate system on the Riemann surface.} Suppose we have a correlation function including $p$ operators $\mathcal{O}_1,...\mathcal{O}_p$, in addition to the $n$ vertex operators representing the external states. We remove a disk from the Riemann surface containing the location of all operators $\mathcal{O}_1,...,\mathcal{O}_p$, but no vertex operators. We fix a coordinate system on this disk denoted $y$ with $|y|<1$, so that each operator $\mathcal{O}_i$ has a corresponding position $y^i$ in this coordinate system. For short we write the complete set of operator insertions as 
\begin{equation}
\mathcal{O}^p = \mathcal{O}_1(y^1)...\mathcal{O}_p(y^p),
\end{equation}
where the upper index $p$ indicates the number of operator insertions. We build the remaining part of the Riemann surface by removing $n$ disks around the punctures, covered by coordinates $w_1,...,w_n$ with $w_a<1$ and $w_a=0$ corresponding to the location of the punctures. Including the disk $y$, the surface now has $n+1$ holes; we cut what remains into $2g+n-1$ components with the topology of a sphere with three holes, and introduce coordinates $z_i$ on these components. The Riemann surface may be reconstructed by specifying transition functions between coordinates identified by gluing:
\begin{eqnarray}
z_i\lineup = f_{ij}(z_j)\label{eq:ztransp}\\
z_i\lineup = f_a(w_a)\\
z_i\lineup = f(y).\label{eq:ytransp}
\end{eqnarray}
Note that at this level the coordinate $y$ is on the same footing as the coordinates $w_a$, but the coordinate $y$ will play a distinct role in defining the measure. From the transition functions we define a surface state acting on $n+1$ copies of $\mathcal{H}$:
\begin{equation}
\langle \Sigma'|:\mathcal{H}^{\otimes n+1}\to \mathbb{C}.
\end{equation}
We use the prime to indicate that $\langle \Sigma'|$ acts on $n+1$ states, including a state represented by the coordinate~$y$. Assuming that the first copy of $\mathcal{H}$ represents operators inserted in the coordinate $y$, we may then represent correlation functions containing operators $\mathcal{O}_1,...\mathcal{O}_p$ through the $n$-fold bra state
\begin{equation}
\langle \Sigma'|\Big(\mathcal{O}^p|0\rangle\Big)\otimes \mathbb{I}^{\otimes n}.
\end{equation}
Suppose that for every point $m$ in the moduli space we chose transition functions $f_{ij},f_a,f$ building a Riemann surface with moduli $m$. From this we can define a surface state $\langle\Sigma'(m)|$ for every $m\in M$; we write $\langle\Sigma'(m)|$ simply as $\langle\Sigma'|$, leaving the dependence on $m$ implicit.
We assume that the transition functions have been defined so that the coordinate $y$ covers all parts of each Riemann surface where we care to insert $\mathcal{O}_1,...,\mathcal{O}_p$. Note that the moduli space carries information about the location of the $n$ punctures represented by the coordinates $w_1,...,w_n$, but does not carry information about the coordinate $y$. We introduce a collection of $n+1$ Schiffer vector fields $v_\mu(m,y)$ and $v^{(a)}_\mu(m,w_a)$ defined so that the analogue of \eq{bosid1}-\eq{bosid3} hold: 
\begin{eqnarray}
d\langle \Sigma'|\lineup =-\langle \Sigma|T'\label{eq:supid1}\\
\ [Q,b']\lineup = -T'\\
db'\lineup = \frac{1}{2}[T',b'],\label{eq:supid3}
\end{eqnarray}
where
\begin{equation}
b' \equiv dm^\mu b_\mu'
\end{equation}
and
\begin{eqnarray}\!
b_\mu' \!\lineup \equiv\! \left(\!\oint\!\frac{dy}{2\pi i} v_\mu(m,\! y)b(y)\!\right)\!\otimes\!\mathbb{I}^{\otimes n}\!+\!\mathbb{I}\!\otimes \!\left(\!\oint\!\frac{dw_1}{2\pi i} v_\mu^{(1)}(m,\!w_1)b(w_1)\!\right)\!\otimes\!\mathbb{I}^{\otimes n-1}\!+\!...\!+\!\mathbb{I}^{\otimes n}\!\otimes\! \left(\!\oint\!\frac{dw_n}{2\pi i} v_\mu^{(n)}(m,\!w_n)b(w_n)\!\right)\nonumber\\
\lineup\ \ \ +\!\left(\!\oint\!\frac{d\bar{y}}{2\pi i} \bar{v}_\mu(m,\!\bar{y})\bar{b}(\bar{y})\!\right)\!\otimes\!\mathbb{I}^{\otimes n}\!+\!\mathbb{I}\!\otimes\! \left(\!\oint\!\frac{d\bar{w}_1}{2\pi i} \bar{v}_\mu^{(1)}(m,\!\bar{w}_1)\bar{b}(\bar{w}_1)\!\right)\!\otimes\!\mathbb{I}^{\otimes n-1}\!+\!...\!+\!\mathbb{I}^{\otimes n}\!\otimes\!\left(\! \oint\!\frac{d\bar{w}_n}{2\pi i} \bar{v}_\mu^{(n)}(m,\!\bar{w}_n)\bar{b}(\bar{w}_n)\!\right).\nonumber\\
\end{eqnarray}
With this we define the measure with operator insertions as
\begin{equation}\langle 
\Omega, \mathcal{O}^p| \equiv \langle\Sigma'|e^{b'}\Big(\mathcal{O}^p|0\rangle\Big)\otimes\mathbb{I}^{\otimes n}.\label{eq:Omeasure}\end{equation}
We label the measure according to the operator insertions it contains. 

It is useful to think of the measure as a differential form on a fiber bundle $Y^p$. The base of $Y^p$ consists of the moduli space $M$ of genus $g$ Riemann surfaces with $n$ punctures together with an even or odd spin structure, and $m^\mu$ are coordinates on the base. The fiber at the point $m^\mu$  consists of $p$ copies of the Riemann surface with the corresponding value of the moduli, and $y^1,...,y^p$ are coordinates on the fiber.  We introduce coordinate 1-forms on the fiber $dy^i,d\bar{y}^i$ and define the exterior derivative on $Y^p$:
\begin{equation}
d = dm^\mu\frac{\d}{\d m^\mu} + dy^i\frac{\d}{\d y^i}+d\bar{y}^i\frac{\d}{\d\bar{y}^i}.
\end{equation}
We assume that $dy^i,d\bar{y}^i$ are uniformly Grassmann odd objects which anticommute with each other, the $dm^\mu$s, and Grassmann odd worldsheet operators. Using the identities \eq{supid1}-\eq{supid3}, it is straightforward to show that the generalization of \eq{bos_measure} in the presence of operator insertions takes the form 
\begin{equation}
(-1)^{\mathcal{O}^p}\big\langle\Omega,\mathcal{O}^p\,\big| Q= - d \big\langle\Omega,\mathcal{O}^p\,\big| \,-\, \big\langle\Omega, (Q-d)\mathcal{O}^p\,\big|\phantom{\Bigg)}, \label{eq:OBRSTid}
\end{equation}
where $d$ now includes differentiation along the fiber directions. On the right hand side, $(Q-d)\mathcal{O}^p$ represents a sum of operator insertions
\begin{eqnarray}
(Q-d)\mathcal{O}^p\lineup  = \Big(Q \mathcal{O}_1(y^1)\Big)...\mathcal{O}_p(y^p)+...+(-1)^{\mathcal{O}_1+...+\mathcal{O}_{p-1}}\mathcal{O}_1(y^1)...\Big(Q\mathcal{O}_p(y^p)\Big)\nonumber\\
\lineup\ \ \ -\Big(d\mathcal{O}_1(y^1)\Big)...\mathcal{O}_p(y^p)-...-(-1)^{\mathcal{O}_1+...+\mathcal{O}_{p-1}}\mathcal{O}_1(z^1)...\Big(d\mathcal{O}_p(y^p)\Big),
\end{eqnarray}
and, for example 
\begin{equation}
d\mathcal{O}_1(y^1) = dy^1\d\mathcal{O}_1(y^1) + d\bar{y}^1\bar{\d}\mathcal{O}_1(y^1).
\end{equation}
Also important in our discussion is the property
\begin{equation}(-1)^{\mathcal{O}^p}\langle\Omega,\mathcal{O}^p|\eta = -\langle\Omega,\eta\mathcal{O}^p|,\end{equation}
where $\eta\mathcal{O}^p$ represents a sum of operator insertions
\begin{equation}\eta\mathcal{O}^p = \Big(\eta \mathcal{O}_1(y^1)\Big)...\mathcal{O}_p(y^p)+...+(-1)^{\mathcal{O}_1+...+\mathcal{O}_{p-1}}\mathcal{O}_1(y^1)...\Big(\eta\mathcal{O}_p(y^p)\Big).
\end{equation}
This is nonzero only if $\mathcal{O}^p$ contains some operators in the large Hilbert space. 

The measure which is relevant for computing superstring scattering amplitudes in the PCO formalism is
\begin{equation}\langle \Omega,X^p|,\label{eq:supermeasure}\end{equation}
where $X^p$ refers to a collection of operator insertions of the form 
\begin{equation}X^p \equiv \Big[X(y^1)-d\xi(y^1)\Big]\ ...\ \Big[X(y^p)-d\xi(y^p)\Big],\label{eq:supermeasureop}\end{equation}
and $X(y)=Q\xi(y)$ is a picture changing operator. If the number of insertions $p$ is chosen appropriately, we obtain nonvanishing correlation functions with Neveu-Schwarz and Ramond external states at the standard pictures $-1$ and $-1/2$. The measure is defined in the small Hilbert space:
\begin{equation}\langle \Omega, X^p| \eta = 0.\end{equation}
Furthermore, since 
\begin{equation}X(y)-d\xi(y) = (Q-d)\xi(y),\end{equation}
we have the property
\begin{equation}\langle\Omega, X^p|Q = -d\langle \Omega,X^p|.\end{equation}
The second term in \eq{OBRSTid} drops out since $Q-d$ squares to zero. Therefore, the superstring measure produces a total derivative on the fiber bundle $Y^p$ when acting on BRST trivial states. 

Naively, we can define a gauge invariant amplitude by integrating the pullback of the superstring measure on a global section of $Y^p$. The difficulty, however, is in finding a global section of $Y^p$ which avoids spurious singularities in the measure. However, it is always possible to find sections of $Y^p$ which avoid spurious singularities locally. We can then attempt to define the amplitude by summing contributions from local sections on disjoint regions of moduli space which avoid spurious poles. Generally there will be discontinuities in the choice of PCOs between disjoint regions, and the amplitude will require additional contributions---the ``vertical corrections"---to cancel boundary terms between different regions when the amplitude contains BRST trivial states. The vertical corrections can be seen to arise from integrating the superstring measure ``along the fiber" at junctions between different regions of the moduli space so as to join local sections into a closed integration cycle in $Y^p$. This is {\it vertical integration}. Next we describe the algebraic approach to the PCO formalism, where the origin of vertical corrections is somewhat different. 

\section{Algebraic Approach}
\label{sec:alg}

In the algebraic approach outlined in the introduction, PCOs are derived by repeatedly inserting $\xi(y)$ in the measure followed by application of the BRST operator. If the location of $\xi$ is not a continuous function of the moduli, \eq{OBRSTid} implies that the BRST operator produces boundary terms from the integration over moduli space at the locus of discontinuities. These boundary terms are the vertical corrections.

\begin{figure}
\begin{center}
\resizebox{2in}{2in}{\includegraphics{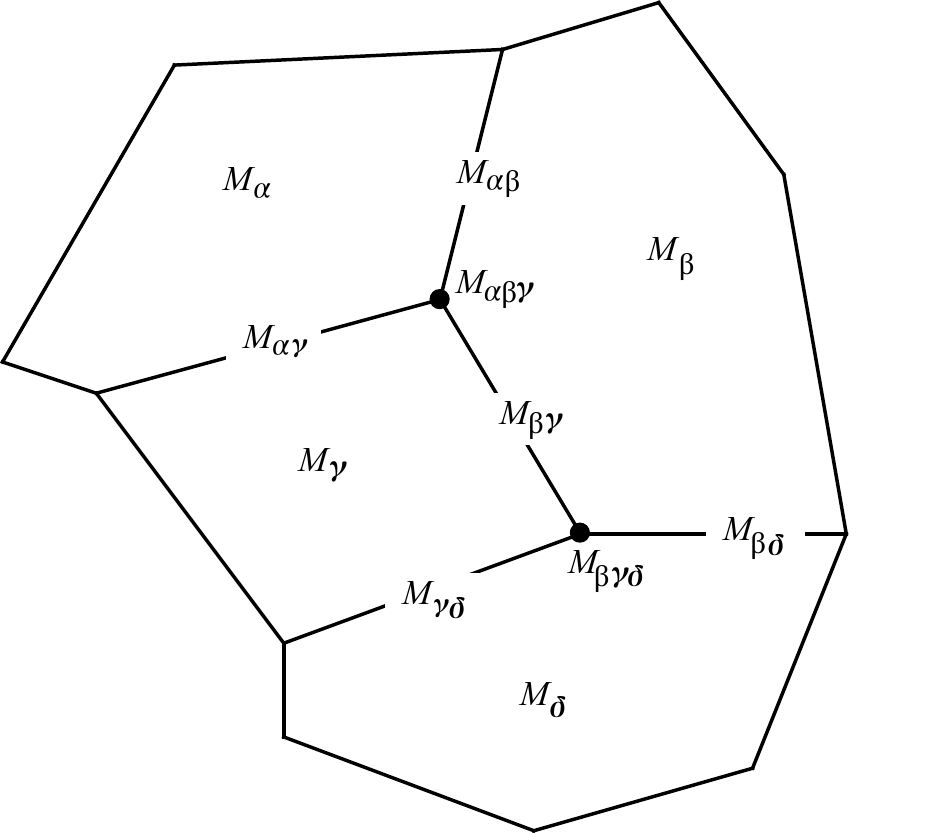}}
\end{center}
\caption{\label{fig:dual} A dual triangulation of (a 2-dimensional) moduli space.}
\end{figure}

We assume that the moduli space is decomposed into regions $M_\alpha$ where the location of $\xi$ varies continuously as a function of the moduli:
\begin{equation}M = \cup_{\alpha}M_\alpha.\end{equation}
The contribution to the amplitude from $M_\alpha$ will turn out to be the pullback of the superstring measure on a local section of $Y^p$ defined on $M_\alpha$. To connect with the discussion of \cite{SenWitten}, we assume that the regions $M_\alpha$ form closed polyhedra which are glued along their faces in such a way as to define a dual triangulation of $M$. However, it should be clear that the general procedure applies regardless of the choice of decomposition of the moduli space. By definition,  all faces of codimension $k$ in a dual triangulation appear at the junction between $k+1$ distinct polyhedra. We will write $M_{\alpha_0...\alpha_k}$ for the codimension $k$ face at the junction of distinct polyhedra $M_{\alpha_0}...M_{\alpha_k}$, so we have
\begin{eqnarray}
\mathrm{codimension}\ 0:\lineup \ \ \ \ M_\alpha\nonumber\\
\mathrm{codimension}\ 1:\lineup\ \ \ \ M_{\alpha\beta} = M_\alpha\cap M_\beta,\ \ \ \alpha,\beta\ \mathrm{distinct}\nonumber\\
\mathrm{codimension}\ 2:\lineup\ \ \ \ M_{\alpha\beta\gamma} = M_\alpha\cap M_\beta\cap M_\gamma\ \ \ \  \alpha,\beta,\gamma\ \mathrm{distinct}\nonumber\\
\vdots\ \ \ \ \ \ \ \ \ \ \lineup\ \ \ \ \ \ \ \ \ \ \ \ \ \ \vdots\ \ .
\end{eqnarray}
See figure \ref{fig:dual}. If the intersection of the polyhedra $M_{\alpha_0},...,M_{\alpha_k}$ is empty, we assume that $M_{\alpha_0...\alpha_k}$ is the empty set. The faces $M_{\alpha\beta}$ and $M_{\beta\alpha}$ are equal as sets, but it is useful to consider them as having opposite orientations as integration cycles in the moduli space. More generally, we assume that 
\begin{equation}\int_{M_{...\alpha_i...\alpha_j...}} = -\int_{M_{...\alpha_j...\alpha_i...}}.\end{equation}
In this sense,  $M_{\alpha_0...\alpha_k}$ is totally antisymmetric in the indices $\alpha_0...\alpha_k$. In particular, $M_{\alpha_0...\alpha_k}$ is the empty set if any two indices are equal. Fixing an orientation on the moduli space induces an orientation on the polyhedra, and the orientation of the higher codimension faces will be determined by 
\begin{equation}\int_{ \d M_{\alpha_0...\alpha_k}} = -\sum_\beta \int_{M_{\alpha_0...\alpha_k\beta}}.
\end{equation}
In this setup we can formulate a useful version of Stokes' theorem. Suppose on each codimension $k$ face $M_{\alpha_0...\alpha_k}$ we have a differential form $\omega_{\alpha_0...\alpha_k}$ which is antisymmetric in the indices $\alpha_0...\alpha_k$. Stokes' theorem implies
\begin{equation}
\frac{1}{(k+1)!}\sum_{\alpha_0...\alpha_k}\int_{M_{\alpha_0...\alpha_k}} d\omega_{\alpha_0...\alpha_k} 
= \frac{1}{(k+2)!}\sum_{\alpha_0...\alpha_{k+1}}\int_{M_{\alpha_0...\alpha_{k+1}}}(\delta\omega)_{\alpha_0...\alpha_{k+1}}. \label{eq:stokes}
\end{equation}
We introduce an operation $\delta$, which acts on an object with antisymmetric indices $\alpha_0...\alpha_k$ to produce an object with antisymmetric indices $\alpha_0...\alpha_{k+1}$. It is defined as
\begin{equation}(\delta\omega)_{\alpha_0...\alpha_{k+1}} = \sum_{n=0}^{k+1}(-1)^n\omega_{\alpha_0...\widehat{\alpha}_n...\alpha_{k+1}},
\end{equation}
where the hat over the index indicates omission. The operation $\delta$ is nilpotent,
\begin{equation}\delta^2=0,\end{equation}
and is related to the \v{C}ech coboundary operator.

\subsection{The Construction}

We propose to express the amplitude in the form\footnote{Since $M$ refers to only one connected component of the moduli space of Riemann surfaces with spin structure, technically $\langle A^p|$ only gives the contribution to the total amplitude coming from either the even or the odd spin structures. Since the location of spurious poles depends on the spin structure, in general we must adjust the choice of dual triangulation, local sections, and vertical corrections separately for the even and odd spin structures. The complete amplitude is then given by adding these contributions.}
\begin{equation}
\langle A^p|=\sum_\alpha \int_{M_\alpha} \langle
\Omega, X^p_\alpha| + \frac{1}{2!}\sum_{\alpha\beta}\int_{M_{\alpha\beta}}\langle\Omega, X^p_{\alpha\beta}|+\frac{1}{3!}\sum_{\alpha\beta\gamma}\int_{M_{\alpha\beta\gamma}}\langle\Omega, X^p_{\alpha\beta\gamma}| + ...\ .\label{eq:amp}
\end{equation}
The first term is the contribution to the amplitude from the pullback of the superstring measure \eq{supermeasure} onto local sections of $Y^p$ on each polyhedron. The operator insertions in the first term are given by
\begin{equation}
X_\alpha^p = \Big[X(y^1_\alpha(m)) - d\xi(y^1_\alpha(m))\Big]\ .\,.\,.\ \Big[X(y^p_\alpha(m)) - d\xi(y^p_\alpha(m))\Big],
\end{equation}
where the points $y^1_\alpha(m),...,y^p_\alpha(m)$ parameterize the location of the PCOs as a function of $m\in M_\alpha$, and characterize the local section of $Y^p$.  The remaining terms in the amplitude are the vertical corrections, and can be arranged hierarchically according to the codimension of the faces in the dual triangulation. The vertical corrections are defined by integrating a measure $\langle\Omega, X_{\alpha_0...\alpha_k}^p|$ over the face $M_{\alpha_0...\alpha_k}$ of the dual triangulation, where $X_{\alpha_0...\alpha_k}^p$ denotes a collection of $p$ operator insertions whose positions are prescribed functions of $m\in M_{\alpha_0...\alpha_k}$. The insertions $X_{\alpha_0...\alpha_k}^p$ are defined to be antisymmetric in the indices $\alpha_0...\alpha_k$, and for even (odd) codimension the insertions are  Grassmann even (odd). Generally, $X_{\alpha_0...\alpha_k}^p$ will be expressed through combinations of $X(y),\d \xi(y)$ and $\xi(y)$, and the goal of the present discussion is to determine what form the insertions take.

The central condition characterizing the vertical corrections is that they lead to a gauge invariant amplitude. From \eq{OBRSTid} we know that
\begin{equation}(-1)^k \langle\Omega, X_{\alpha_0...\alpha_k}^p| Q =  -d \langle\Omega, X_{\alpha_0...\alpha_k}^p| -\big\langle \Omega,(Q-d)X_{\alpha_0...\alpha_k}^p\big|.\end{equation}
Using Stokes' theorem \eq{stokes}, gauge invariance implies that the operator insertions $X^p_{\alpha_0...\alpha_k}$ satisfy 
\begin{equation}(Q-d)X^p_{\alpha_0...\alpha_k}-(\delta X^p)_{\alpha_0...\alpha_k}=0\phantom{\Bigg(} .\label{eq:gaugeX}\end{equation}
The operator $\delta$ acts on the insertions $X^p_{\alpha_0...\alpha_{k-1}}$ corresponding to the faces of one fewer codimension:
\begin{equation}(\delta X^p)_{\alpha_0...\alpha_k} = \sum_{n=0}^k(-1)^n X_{\alpha_0...\widehat{\alpha}_n...\alpha_k}^p.\end{equation}
All terms in \eq{gaugeX} are evaluated at a common point $m\in M_{\alpha_0...\alpha_k}$. 

To solve \eq{gaugeX}, we propose that the physical amplitude can be expressed as the BRST variation of  a {\it gauge amplitude}: 
\begin{equation}\langle A^p| = \langle\alpha^p| Q.\end{equation}
The gauge amplitude $\langle \alpha^p|$ is expressed in a form analogous to \eq{amp}:
\begin{equation}
\langle \alpha^p|=\sum_\alpha \int_{M_\alpha} \langle\Omega, \Xi^p_\alpha| - \frac{1}{2!}\sum_{\alpha\beta}\int_{M_{\alpha\beta}}\langle\Omega, \Xi^p_{\alpha\beta}|+\frac{1}{3!}\sum_{\alpha\beta\gamma}\int_{M_{\alpha\beta\gamma}}\langle \Omega,\Xi^p_{\alpha\beta\gamma}| - ...\ .\label{eq:gaugeamp}
\end{equation}
For convenience, we take the signs in this series to alternate. On each face of the dual triangulation we have a measure defined by a collection of $p$ operator insertions $\Xi^p_{\alpha_0...\alpha_k}$. The insertions $\Xi_{\alpha_0...\alpha_k}^p$ are antisymmetric in the indices $\alpha_0...\alpha_k$, and for even (odd) codimension they are Grassmann odd (even). Typically, the insertions $\Xi_{\alpha_0...\alpha_k}^p$ depend on the zero mode of the $\xi$ ghost. Taking the BRST variation of the gauge amplitude gives a formula for the insertions $X_{\alpha_0...\alpha_k}^p$: 
\begin{equation}X_{\alpha_0...\alpha_k}^p = (Q-d)\Xi_{\alpha_0...\alpha_k}^p+(\delta \Xi^p)_{\alpha_0...\alpha_k}\phantom{\Bigg(}.\label{eq:gaugexi}\end{equation}
The operator $\delta$ acts on the insertions $\Xi^p_{\alpha_0...\alpha_{k-1}}$ corresponding to the faces of one fewer codimension,
\begin{equation}(\delta \Xi^p)_{\alpha_0...\alpha_k} = \sum_{n=0}^k(-1)^n \Xi_{\alpha_0...\widehat{\alpha}_n...\alpha_k}^p,\end{equation}
and all terms in \eq{gaugexi} are evaluated at a common point on $M_{\alpha_0...\alpha_k}$.  Note that, schematically, gauge invariance requires that $X^p_{\alpha_0...\alpha_k}$ is annihilated by $Q-d-\delta$, and this follows from \eq{gaugexi} because 
\begin{equation}(Q-d-\delta)(Q-d+\delta) = (Q-d)^2-\delta^2 = 0.\end{equation}
Since the physical amplitude is defined in the small Hilbert space, we know that the insertions $X_{\alpha_0...\alpha_k}^p$ must be independent of the $\xi$ zero mode:
\begin{equation}\eta X_{\alpha_0...\alpha_k}^p = 0.\end{equation}
From \eq{gaugexi}, we therefore learn that $\eta\Xi_{\alpha_0...\alpha_k}^p$ satisfies
\begin{equation}
(Q-d)\eta\Xi_{\alpha_0...\alpha_k}^p - (\delta\eta\Xi^p)_{\alpha_0...\alpha_k} = 0.
\end{equation}
Interestingly, this implies that the operator insertions given by $\eta\Xi_{\alpha_0...\alpha_k}^p$ define a gauge invariant amplitude. Since $\eta$ carries picture $-1$, it is natural to interpret $\eta\Xi_{\alpha_0...\alpha_k}^p$ as defining an amplitude with one fewer PCO insertion:
\begin{equation}\eta\Xi_{\alpha_0...\alpha_k}^p = X_{\alpha_0...\alpha_k}^{p-1}.\end{equation}
Thus we have the relation
\begin{equation}\langle \alpha^p|\eta = \langle A^{p-1}|,\end{equation}
where $\langle A^{p-1}|$ is defined by insertions $X_{\alpha_0...\alpha_k}^{p-1}$. We can apply this procedure again, relating $\langle A^{p-1}|$ to the amplitude $\langle A^{p-2}|$ containing two fewer PCO insertions, and continue all the way down until we have the amplitude $\langle A^0|$ where PCOs are absent. 

This leads to the following procedure for deriving gauge invariant amplitudes. The ``insertions" defining an amplitude without PCOs can be trivially written 
\begin{equation}
X_\alpha^0 = 1,\ \ \ \ \ \ X_{\alpha_0...\alpha_k}^0 = 0\ \ \ (k\geq 1) .\label{eq:X0}
\end{equation}
The second equation says that there are no vertical corrections in the absence of PCOs. Since $X^0_{\alpha_0...\alpha_k}$ is independent of the $\xi$ zero mode, it can be expressed in $\eta$-exact form:
\begin{equation}
X^0_{\alpha_0...\alpha_k} = \eta\Xi^1_{\alpha_0...\alpha_k}.
\end{equation}
The expression for $\Xi^1_{\alpha_0...\alpha_k}$ is not unique, but let us assume that we have made some choice. We can then plug into \eq{gaugexi} to derive an expression for the insertions $X^1_{\alpha_0...\alpha_k}$ defining the amplitude with a single PCO. By construction, $X^1_{\alpha_0...\alpha_k}$ will be independent of the $\xi$ zero mode and can be expressed in $\eta$-exact form:
\begin{equation}
X^1_{\alpha_0...\alpha_k} = \eta\Xi^2_{\alpha_0...\alpha_k}.
\end{equation}
Substituting into \eq{gaugexi} gives the insertions $X_{\alpha_0...\alpha_k}^2$ defining the amplitude with two PCOs. Continuing this process for $p$ steps we arrive at the insertions $X_{\alpha_0...\alpha_k}^p$, as desired.

The solution generated by this procedure is not unique. For most purposes it does not matter how the solution is chosen as long as the PCO insertions in the final amplitude avoid spurious poles. As we will demonstrate later, Sen and Witten give a class of solutions for the vertical corrections which can be generated by this procedure, but not the most general solution.

\subsection{Examples}
\label{subsec:examples}

Let us give some examples to see what the vertical corrections look like. Consider first an amplitude containing one PCO. We must find a set of insertions $\Xi^1_{\alpha_0...\alpha_k}$ satisfying 
\begin{equation}X_{\alpha_0...\alpha_k}^0 = \eta\Xi_{\alpha_0...\alpha_k}^1.\end{equation}
We can choose for example 
\begin{eqnarray}
\Xi_\alpha^1 = \xi(y_\alpha^1(m)),\ \ \ \ \ \ \ \ \ \ \ 
\Xi_{\alpha_0...\alpha_k}^1 =0\ \ \ (k\geq 1),\label{eq:xi1}
\end{eqnarray}
where $y_\alpha^1(m)$ gives the location of a $\xi$ insertion on the Riemann surface as a function of $m\in M_\alpha$ in each polyhedron. We may determine the insertions $X^1_{\alpha_0...\alpha_k}$ by substituting into \eq{gaugexi}:
\begin{eqnarray}
X_\alpha^1 \lineup = (Q-d)\Xi_\alpha^1 \nonumber\\
X_{\alpha\beta}^1\lineup = (Q-d)\Xi_{\alpha\beta}^1+\Xi_\beta^1-\Xi_\alpha^1\nonumber\\
X_{\alpha\beta\gamma}^1\lineup = (Q-d)\Xi_{\alpha\beta\gamma}^1 + \Xi_{\beta\gamma}^1 - \Xi_{\alpha\gamma}^1 + \Xi_{\alpha\beta}^1\nonumber\\
\lineup\vdots\ \ .
\end{eqnarray}
This gives 
\begin{eqnarray}
X_\alpha^1 \lineup = X(y_\alpha^1)-d\xi(y_\alpha^1) \nonumber\\
X_{\alpha\beta}^1\lineup = \xi(y_\beta^1) - \xi(y_\alpha^1)\label{eq:X1ab}\nonumber\\
X_{\alpha\beta\gamma}^1\lineup = 0 \nonumber\\
\lineup\vdots\ \ .
\end{eqnarray}
The vertical corrections on the faces of codimension 2 and higher vanish. Here and in later equations we will not explicitly indicate the dependence of the fiber coordinates $y^i_\alpha$ on the moduli, unless needed for clarity.

As expected, $X_\alpha^1$ is the pullback of the superstring measure \eq{supermeasure} onto a local section of $Y^1$ defined by $y_\alpha^1(m)$. The insertions $X_{\alpha\beta}^1$ have a simple interpretation in terms of vertical integration. Let us make a brief detour to spell out what this means in the current setup. Let $\mathcal{M}_\alpha\subset Y^p$ denote the local section of $Y^p$ defined on each face $M_\alpha$ of the dual triangulation. Let $\mathcal{M}_{\alpha_0...\alpha_k}\subset Y^p$ denote submanifolds of $Y^p$---the ``vertical segments"---which, with a suitable orientation, connect the local sections to form a closed integration cycle in $Y^p$. We assume that the orientation of $\mathcal{M}_{\alpha_0..\alpha_k}$ is antisymmetric in the indices, and postulate that the projection from  $Y^p$ down to the moduli space $M$ maps the vertical segments $\mathcal{M}_{\alpha_0...\alpha_k}$ down to the faces $M_{\alpha_0...\alpha_k}$ of the dual triangulation. This implies that the vertical segments $\mathcal{M}_{\alpha_0...\alpha_k}$ can be parameterized by coordinates on $M_{\alpha_0...\alpha_k}$ together with $k$ coordinates tangent to the fiber. The basic idea is to express the amplitude as
\begin{equation}
\langle A^p| = \sum_{\alpha}\int_{\mathcal{M}_{\alpha}}\langle \Omega,X^p| + \frac{1}{2!}\sum_{\alpha,\beta}\int_{\mathcal{M}_{\alpha\beta}}\langle\Omega,X^p|+\frac{1}{3!}\sum_{\alpha,\beta,\gamma}\int_{\mathcal{M}_{\alpha\beta\gamma}}\langle\Omega,X^p|+...\ ,
\end{equation}
where in each term we take the pullback of the superstring measure \eq{supermeasure} on the corresponding submanifold of $Y^p$. If we integrate out the fiber coordinates on the vertical segments, this gives an expression for the amplitude as postulated in \eq{amp}. We can work this out fairly easily in the case where there is only one PCO. Let us choose a coordinate system on $\mathcal{M}_{\alpha\beta}$ corresponding to coordinates on $M_{\alpha\beta}$ together with an additional coordinate $t\in[0,1]$ parameterizing the fiber direction. The submanifold $\mathcal{M}_{\alpha\beta}$ is defined by specifying the fiber coordinate $y^1$ as a function of $m\in M_{\alpha\beta}$ and $t$. Since $\mathcal{M}_{\alpha\beta}$ must join the local sections $\mathcal{M}_\alpha$ and $\mathcal{M}_\beta$, we require that 
\begin{equation}
y^1(m,t)|_{t=1} = y_\beta^1(m),\ \ \ y^1(m,t)|_{t=0} = y^1_\alpha(m).
\end{equation}
We then find
\begin{eqnarray}
\int_{\mathcal{M}_{\alpha\beta}}\langle \Omega,X^1| \lineup = \int_{M_{\alpha\beta}}\int_t\ \langle \Omega, X(y^1(m,t)) - d\xi(y^1(m,t))|\nonumber\\
\lineup = \int_{M_{\alpha\beta}}\int_0^1dt\frac{d}{dt}\langle \Omega,\xi(y^1(m,t))|\nonumber\\
\lineup = \int_{M_{\alpha\beta}}\langle \Omega,\xi(y^1_\beta(m)) -\xi(y^1_\alpha(m))|\nonumber\\
\lineup =  \int_{M_{\alpha\beta}}\langle \Omega,X^1_{\alpha\beta}|.
\end{eqnarray}
The only part of the measure with the 1-form $dt$ is a total derivative with respect to $t$, and integrating out the fiber coordinate gives \eq{X1ab}. Note that, in this case, the vertical correction only depends on the boundary of $\mathcal{M}_{\alpha\beta}$, not on how $\mathcal{M}_{\alpha\beta}$ is chosen in the interior. This is a special occurrence since we are dealing with only one PCO. With more PCOs, the part of the measure proportional to $dt$ is not a total derivative, and generally the vertical corrections will depend on the choice of vertical segments. This ambiguity corresponds in the algebraic formalism to the different possible ways of expressing an amplitude in $\eta$ exact form.

Let us continue to the case of two PCOs. We must find a set of insertions $\Xi_{\alpha_0...\alpha_k}^2$ satisfying
\begin{equation}X_{\alpha_0...\alpha_k}^1 = \eta\Xi_{\alpha_0...\alpha_k}^2.\end{equation}
We can find a solution by multiplying $X_{\alpha_0...\alpha_k}^1$ by an insertion of $\xi$:
\begin{eqnarray}
\Xi_\alpha^2 \lineup = \xi(y_\alpha^2(m))\Big[X(y_\alpha^1(m))-d\xi(y_\alpha^1(m))\Big]\nonumber\\
\Xi_{\alpha\beta}^2 \lineup = \xi(y^2_{\alpha\beta}(m)) \Big[\xi(y_\beta^1(m)) - \xi(y_\alpha^1(m))\Big]\nonumber\\
\Xi_{\alpha\beta\gamma}^2 \lineup = 0\nonumber\\
\lineup \vdots\ \ .
\end{eqnarray}
Here $y^2_\alpha(m)$ gives the location of a new $\xi$ insertion on the codimension 0 faces as a function of $m\in M_\alpha$, and $y_{\alpha\beta}^2(m)$ gives the location of a $\xi$ insertion on the codimension 1 faces as a function of $m\in M_{\alpha\beta}$. The insertions $X_{\alpha_0...\alpha_k}^2$ are given by substituting into
\begin{eqnarray}
X_\alpha^2 \lineup = (Q-d)\Xi_\alpha^2 \nonumber\\
X_{\alpha\beta}^2\lineup = (Q-d)\Xi_{\alpha\beta}^2+\Xi_\beta^2-\Xi_\alpha^2\nonumber\\
X_{\alpha\beta\gamma}^2\lineup = (Q-d)\Xi_{\alpha\beta\gamma}^2 + \Xi_{\beta\gamma}^2 - \Xi_{\alpha\gamma}^2 + \Xi_{\alpha\beta}^2\nonumber\\
X_{\alpha\beta\gamma\delta}^2\lineup = (Q-d)\Xi_{\alpha\beta\gamma\delta}^2 + \Xi_{\beta\gamma\delta}^2 - \Xi_{\alpha\gamma\delta}^2 +\Xi_{\alpha\beta\delta}^2 -\Xi_{\alpha\beta\gamma}^2\nonumber\\
\lineup\vdots\ \ ,
\end{eqnarray}
which gives
\begin{eqnarray}
X_\alpha^2 \lineup = \Big[X(y_\alpha^2)-d\xi(y_\alpha^2)\Big]\Big[X(y_\alpha^1)-d\xi(y_\alpha^1)\Big] \nonumber\\
X_{\alpha\beta}^2\lineup = \Big[X(y_{\alpha\beta}^2)-d\xi(y_{\alpha\beta}^2)\Big]\Big[\xi(y_\beta^1) - \xi(y_\alpha^1)\Big]\nonumber\\
\lineup\ \ \ +\Big[\xi(y_{\alpha\beta}^2)- \xi(y_\alpha^2)\Big]\Big[X(y_\alpha^1)-d\xi(y_\alpha^1)\Big] -\Big[\xi(y_{\alpha\beta}^2)-\xi(y_\beta^2)\Big]\Big[X(y_\beta^1)-d\xi(y_\beta^1)\Big]  \nonumber\\
X_{\alpha\beta\gamma}^2\lineup = \xi(y^2_{\beta\gamma}) \Big[\xi(y_\gamma^1) - \xi(y_\beta^1)\Big]-\xi(y^2_{\alpha\gamma}) \Big[\xi(y_\gamma^1) - \xi(y_\alpha^1)\Big]+\xi(y^2_{\alpha\beta}) \Big[\xi(y_\beta^1) - \xi(y_\alpha^1)\Big]\nonumber\\
X_{\alpha\beta\gamma\delta}^2\lineup = 0\nonumber\\
\lineup\vdots\ \ .\label{eq:X2}
\end{eqnarray}
The vertical corrections on the faces of codimension 3 and higher vanish. As expected, $X_\alpha^2$ is the pullback of the superstring measure \eq{supermeasure} onto a local section of $Y^2$ defined by $y^1_\alpha(m),y^2_\alpha(m)$. However, it is not immediately obvious how the higher corrections should be interpreted through vertical integration. In fact, for generic choice of $y_{\alpha\beta}^2(m)$, the vertical corrections are already outside the class which can be obtained from the Sen-Witten procedure. Another thing to mention, as can be seen in the expression for $X^2_{\alpha\beta\gamma}$, is that the operator insertions appearing in the vertical corrections are not directly expressed as differences of $\xi$s. Since the insertions are together independent of the $\xi$ zero mode, it is possible to express the vertical corrections in terms of differences of $\xi$s, but there does not seem to be a preferred way to do this. This means that there is some ambiguity in the interpretation of the vertical corrections in terms of operators in the small Hilbert space. The vertical corrections of Sen and Witten, however, are canonically presented using differences of $\xi$s. This reflects the fact that their construction is intrinsically formulated in the small Hilbert space, which is a notable difference from the algebraic approach, and for some purposes may be an advantage.

\subsection{Dependence on the Choice of PCOs}

The construction of the amplitude requires a lot of data: a choice of dual triangulation, local sections on the polyhedra, and vertical corrections on the higher codimension faces. The final result for the on-shell amplitude, however, should be independent of these choices. One advantage of the algebraic formalism is that this is fairly easy to see, as we now describe.

Suppose we have amplitudes $\langle A^p|$ and $\langle B^p|$ constructed following the algebraic procedure we have described. Associated with these is a hierarchy of amplitudes and gauge amplitudes with an intermediate number of PCOs:
\begin{eqnarray}
\lineup \langle A^0| \ \xleftarrow{\ \eta\ }\langle\ \alpha^1|\ \xrightarrow{\ Q\ }\ \langle A^1|\ \xleftarrow{\ \eta\ }\ \ ...\ \ \langle \alpha^p|\ \xrightarrow{\ Q\ }\ \langle A^p|\nonumber\\
\lineup \langle B^0|\  \xleftarrow{\ \eta\ }\ \langle \beta^1|\ \xrightarrow{\ Q\ }\ \langle B^1|\ \xleftarrow{\ \eta\ }\ \ ...\ \ \langle\beta^p|\ \xrightarrow{\ Q\ }\ \langle B^p|.
\end{eqnarray}
The amplitudes $\langle A^p|$ and $\langle B^p|$ are physically equivalent if we have the relation
\begin{equation}\langle A^p| - \langle B^p| = \langle \Lambda^p|Q\end{equation}
for some $\langle \Lambda^p|$ in the small Hilbert space. By the nature of the construction of the amplitudes, we may find a solution for $\langle \Lambda^p|$ in the form
\begin{equation}\langle \Lambda^p| = \langle \alpha^p| - \langle \beta^p| +\langle D^p|Q\end{equation}
where $\langle D^p|$ must be chosen so that $\langle \Lambda^p|$ is in the small Hilbert space. This implies
\begin{eqnarray}
0\lineup = \langle \Lambda^p|\eta\nonumber\\
\lineup = \langle A^{p-1}| - \langle B^{p-1}| -\langle D^p|\eta Q\nonumber\\
\lineup = \Big(\langle \Lambda^{p-1}|-\langle D^p| \eta\Big)Q,
\end{eqnarray}
where we assume that the difference in $\langle A^{p-1}|$ and $\langle B^{p-1}|$ can be expressed as the BRST variation of some $\langle\Lambda^{p-1}|$ in the small Hilbert space. Therefore $\langle D^p|$ can be determined by solution of the equation
\begin{equation}\langle D^p|\eta = \langle\Lambda^{p-1}|.\end{equation}
Now in a similar way we can look for $\langle \Lambda^{p-1}|$ in the form
\begin{equation}\langle \Lambda^{p-1}| = \langle \alpha^{p-1}|-\langle\beta^{p-1}| +\langle D^{p-1}|Q,\end{equation}
and by the same argument as above, we find that $\langle D^{p-1}|$ solves
\begin{equation}\langle D^{p-1}|\eta = \langle \Lambda^{p-2}|,\end{equation}
where $\langle \Lambda^{p-2}|$ is in the small Hilbert space and its BRST variation computes the difference in $\langle A^{p-2}|$ and $\langle B^{p-2}|$. Continuing this way we find that the state $\langle \Lambda^q|$ with $1\leq q\leq p$ can be determined from the state $\langle \Lambda^{q-1}|$ corresponding to one fewer PCO. If we assume that the  amplitudes without PCOs are identical,\footnote{It is possible that the amplitudes $\langle A^0|$ and $\langle B^0|$ can differ off-shell, for example if the transition functions defining the surface states for each value of the moduli differ. In this case, there should be a nonvanishing $\langle \Lambda_0|$ such that $\langle A^0|-\langle B^0|=\langle\Lambda_0|Q$, and we can build $\langle \Lambda^p|$ starting from there. However, here we do not address the construction of $\langle \Lambda_0|$.}
\begin{equation}\langle A^0|=\langle B^0|,\end{equation}
we can take
\begin{equation}\langle \Lambda^0|=0.\end{equation}
From this starting point we may then find a solution for all states $\langle \Lambda^q|$ with $1\leq q\leq p$. Assuming this is done in such a way that $\langle \Lambda^p|$ avoids spurious poles, this proves that the amplitudes $\langle A^p|$ and $\langle B^p|$ are physically equivalent.

\section{Sen-Witten Approach}
\label{sec:SenWitten}

In this section we describe the Sen-Witten solution for the vertical corrections. Before entering into technicalities, let us take a moment to motivate the origin of the structure. A key point is that the measure \eq{supermeasure} is not a total derivative in the fiber coordinates. This means that integration along a typical vertical segment at the interface between two polyhedra will produce  an integral over a 1-parameter family of PCO configurations whose positions continuously interpolate between those prescribed by neighboring local sections. However, at some point along the vertical segment the configuration of PCO positions is expected to encounter a spurious pole---if this wasn't the case, the two local sections could be smoothly deformed and joined into a larger section. When the amplitude requires only one PCO, there is a way out: the measure is a total derivative along the fiber, and vertical integration produces a finite difference between $\xi$ insertions at different positions, with no integration of PCOs in between. Thus we may use vertical integration to ``jump" across spurious poles. This mechanism can be generalized for two or more PCOs, but with vertical segments of a particular kind. The idea is to form a path in the fiber consisting of $p$ segments, with $p$ the number of PCOs. On each segment we move one PCO from its initial to its final position, while keeping the position of the other PCOs fixed. On each segment the measure is a total derivative in the fiber coordinates, and vertical integration produces a finite difference of $\xi$ insertions between different positions. The upshot is the following: Suppose the polyhedra $M_{\alpha_0}$ and $M_{\alpha_1}$ share a common boundary $M_{\alpha_0\alpha_1}$, and the two polyhedra come with local sections of $Y^p$  characterized respectively by PCO positions $y^1_{\alpha_0},...,y^p_{\alpha_0}$ and $y^1_{\alpha_1},...,y^p_{\alpha_1}$. Then vertical integration along the class of segments described above will produce sums of PCOs at a discrete set of positions
\begin{equation}y^1_{\alpha_{N^1}},...,y^p_{\alpha_{N^p}},\end{equation}
where $N^1,...,N^p$ take values of $0$ or $1$. In particular, there is no integration over a continuous family of PCO positions connecting $y^1_{\alpha_0},...,y^p_{\alpha_0}$ and $y^1_{\alpha_1},...,y^p_{\alpha_1}$, and we can use this fact to ``jump" across spurious poles. Note that the integers $(N^1,...,N^p)$ can be interpreted as vertices of a $p$-dimensional cube. This leads to a natural connection between the Sen-Witten vertical corrections and $p$-dimensional lattices, which we further develop in the language of {\it difference forms}. 

\subsection{Difference Forms}

The Sen-Witten vertical corrections can be naturally expressed using an analogue of differential forms on a lattice, called {\it difference forms}. The concept is fairly straightforward, but it is necessary present the definitions. A more general presentation can be found in \cite{difference}.

We consider a cubic lattice in $\mathbb{R}^p$ given by the set of points 
\begin{equation}(N^1,N^2,...,N^p),\end{equation}
where $N^i$ are integers in the range $0\leq N^i\leq k$ for some $k$. We introduce shift vectors between the lattice sites,
\begin{equation}e_i\equiv(\underbrace{0,\ ...\ ,0}_{i-1\ \mathrm{times}},1,0,\ ...\ ,0),\end{equation}
so that any point in the lattice can be written at $\vec{N}=N^i e_i$.

\begin{figure}
\begin{center}
\resizebox{3in}{2.6in}{\includegraphics{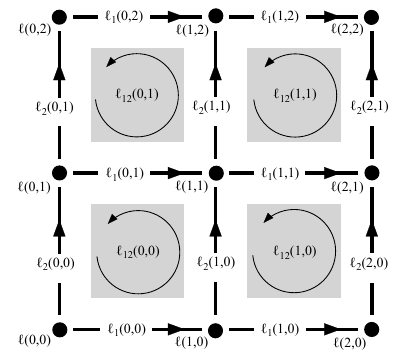}}
\end{center}
\caption{\label{fig:links} Links forming a $2\times 2$ lattice. The orientation on the links is indicated by the arrows.}
\end{figure}

We consider a chain complex given by formal sums of faces on the lattice, which we call {\it links}, together with a naturally defined boundary operator.  For each point $\vec{N}$ in the lattice we introduce a {\it 0-link} denoted $\ell(\vec{N})$. A {\it lattice 0-chain} is defined as a formal sum of 0-links with integer coefficients,
\begin{equation}
C_0 = \sum_{\vec{N}} c(\vec{N})\ell(\vec{N}),
\end{equation}
for $c(\vec{N})\in \mathbb{Z}$. For each line segment connecting neighboring lattice points $\vec{N},\vec{N}+e_i$, we introduce a {\it 1-link} denoted $\ell_i(\vec{N})$. A {\it lattice 1-chain} is defined as a formal sum of 1-links with integer coefficients,
\begin{equation}
C_1 =\sum_{\vec{N}}c^i(\vec{N})\ell_i(\vec{N}),
\end{equation}
for $c^i(\vec{N})\in \mathbb{Z}$. Repeated indices $i$ will always be summed over allowed values; generically, this corresponds to all shift vectors $e_i$, but at the edges of the lattice, it can happen that a shift vector would exit the lattice---that is, while $\vec{N}$ is in the lattice, $\vec{N}+e_i$ is not. Such values of $i$ will always be excluded from sums. Generally, for each $h$-dimensional cube on the lattice with corners at $\vec{N},\vec{N}+e_{i_1},...,\vec{N}+e_{i_h}$ we introduce an {\it h-link} denoted $\ell_{i_1...i_h}(\vec{N})$. We take this to be antisymmetric in the indices:
\begin{equation}\ell_{...i...j...}(\vec{N}) =-\ell_{...j...i...}(\vec{N}).\end{equation}
A {\it lattice h-chain} is given by formal sums of $h$-links with integer coefficients,
\begin{equation}C_h=\frac{1}{h!}\sum_{\vec{N}} c^{i_1...i_h}(\vec{N})\ell_{i_1...i_h}(\vec{N}),\end{equation}
with $c^{i_1...i_h}(\vec{N})$ a set of integers, antisymmetric in $i_1...i_h$. See figure \ref{fig:links}. We introduce a boundary operator defined
\begin{equation}\d \ell_{i_1...i_h}(\vec{N}) = \sum_{n=1}^h(-1)^{n+1}\Big(\ell_{i_1...\widehat{i}_n...i_h}(\vec{N}+e_{i_n}) - \ell_{i_1...\widehat{i}_n...i_h}(\vec{N})\Big),\label{eq:bdry}\end{equation}
where the hat indicates omission. For example, the boundary of a $1$-link is given by the difference of $0$-links at either end:
\begin{equation}\d \ell_i(\vec{N}) = \ell(\vec{N}+e_i)-\ell(\vec{N}).\end{equation}
The boundary operator is nilpotent.

Next we introduce a collection of objects called {\it dual links}. For every $h$-link $\ell_{i_1...i_h}(\vec{N})$, we introduce a corresponding {\it dual h-link}, denoted  $d^{i_1...i_h}(\vec{N})$. This is a linear map from lattice chains into numbers, defined by 
\begin{equation}
d^{i_1...i_h}(\vec{N})\Big[\ell_{j_1...j_{h'}}(\vec{N}')\Big] = \delta_{\vec{N}\vec{N}'}\delta_{hh'}\delta^{i_1}_{[j_1}...\delta^{i_h}_{j_h]},
\end{equation}
which is antisymmetric in the indices $i_1...i_h$. A {\it difference form}, or more specifically a difference $h$-form, is a linear combination of dual $h$-links,
\begin{equation}
\omega_h = \frac{1}{h!}\sum_{\vec{N}} \omega_{i_1...i_h}(\vec{N})d^{i_1...i_h}(\vec{N}),
\end{equation}
where the coefficients $\omega_{i_1...i_h}(\vec{N})$ are a collection of quantities (not necessarily integers) which are antisymmetric in $i_1...i_h$. The action of a difference form $\omega$ on a lattice chain $C$ will be written
\begin{equation}
\omega[C].
\end{equation}
This can be understood as a lattice analogue of integration of a differential form over a manifold. We may also introduce a notion of wedge product between difference forms,
\begin{equation}
d^{i_1...i_n}(\vec{N})\wedge d^{j_1...j_m}(\vec{N}') = \delta_{\vec{N}\vec{N}'}d^{i_1...i_n j_1...j_m}(N),\label{eq:wedge}
\end{equation}
and in this way define an exterior algebra. We usually drop the wedge symbol when multiplying difference forms. We assume that dual $h$-links are uniformly Grassmann even/odd objects for $h$ even/odd, and correspondingly commute or anticommute through Grassmann even or odd worldsheet operators and differential forms. Consistency implies that lattice $h$-chains must also be uniformly Grassmann even/odd for $h$ even/odd.

Let $f(\vec{N})$ be collection of numbers associated to each point on the lattice. If $\vec{N}$ and $\vec{N}+e_i$ are both in the lattice, we can define the {\it difference operator}
\begin{equation}\Delta_i f(\vec{N}) \equiv f(\vec{N}+e_i)-f(\vec{N}).\end{equation}
The {\it exterior difference operator} is defined
\begin{equation}\Delta = \sum_{\vec{N}} d^i(\vec{N})\Delta_i.\end{equation}
In this formula, $\Delta_i$ is assumed to act on coefficients multiplying the dual links, but not the dual links themselves. The exterior difference operator is nilpotent,
\begin{equation}\Delta^2=0,\end{equation}
but is {\it not} a derivation of the wedge product. If $\omega$ is a difference form and $C$ a lattice chain, we have the identity 
\begin{equation}\Delta\omega[C] = (-1)^{\omega-C}\omega[\d C].\phantom{\Bigg]}\end{equation}
This is the analogue of Stokes' theorem on the lattice. The sign refers to the difference in Grassmannality between $\omega$ and $C$; this will only be nonzero if the coefficients of $\omega$ are anticommuting quantities.

\subsection{The Construction}

The construction of the Sen-Witten vertical corrections goes as follows: For each codimension $k$ face of the dual triangulation of moduli space, we associate a $p$-dimensional cubic lattice with $(k+1)^p$ lattice sites, where $p$ is the number of PCOs. On this lattice there is a natural analogue of the superstring measure, which we call the {\it lattice measure}, expressed using difference forms. The vertical corrections are given by acting this measure on an appropriately chosen lattice $k$-chain.

We assume that on each polyhedron $M_\alpha$ we have a choice of local section of $Y^p$ describing the location of $p$ PCOs:
\begin{equation}(y^1_\alpha,\,y^2_\alpha,\,...,\,y^p_\alpha).\end{equation}
We leave the dependence on $m\in M_\alpha$ implicit. It is convenient to specify a total ordering on the collection of polyhedra, so for two distinct polyhedra we have either $M_\alpha<M_\beta$ or $M_\beta<M_\alpha$. Any codimension $k$ face of the dual triangulation can be written as $M_{\alpha_0...\alpha_k}$, where the indices are ordered so that $M_{\alpha_0}<...<M_{\alpha_k}$. We then associate with $M_{\alpha_0...\alpha_k}$ a $p$-dimensional cubic lattice consisting of points
\begin{equation}(N^1,N^2,...,N^p)\end{equation}
with $N^i$ integers satisfying $0\leq N^i\leq k$. Each point in the lattice corresponds to a collection of $p$ PCO insertions on the Riemann surface with coordinates given by
\begin{equation}(y^1_{\alpha_{N^1}},y^2_{\alpha_{N^2}},...,y^p_{\alpha_{N^p}}).\label{eq:coord}\end{equation}
All coordinates in \eq{coord} are evaluated on a common point $m\in M_{\alpha_0...\alpha_k}$. The total ordering of polyhedra allows us to associate to each integer value of the $i$th coordinate $N^i$ a polyhedron which intersects the face of the dual triangulation. This then determines which local section gives the position of the $i$th PCO.

Next we define a generalization of the superstring measure associated to the lattice. In \eq{supermeasureop}, the $i$th PCO contributes to the superstring measure through the factor:
\begin{equation}X(y^i)-d\xi(y^i).\label{eq:supmeas2}\end{equation}
This will effectively be generalized by replacing $d\xi$ with $(d+\Delta)\xi$, where $d$ is the exterior derivative on $M_{\alpha_0...\alpha_k}$ and $\Delta$ is the exterior difference operator on the associated lattice. Specifically, the $i$th PCO contributes through the factor
\begin{equation}
\Big(X(y^i_{\alpha_0})-(d+\Delta)\xi(y^i_{\alpha_0})\Big)\sum_{\vec{N},N^i=0} d(\vec{N}) + ...+\Big(X(y^i_{\alpha_k})-(d+\Delta)\xi(y^i_{\alpha_k})\Big)\sum_{\vec{N},N^i=k} d(\vec{N}),\label{eq:gensupermeas2}
\end{equation}
where
\begin{equation}\sum_{\vec{N},N^i=n}\end{equation}
denotes the sum over a codimension 1 plane in the lattice consisting of all points which share a common $i$th coordinate $N^i$ equal to $n$. The fact that sums appear in \eq{gensupermeas2}, and not in \eq{supmeas2}, is a feature of the notation; in fact they are precisely analogous. For difference forms, every point on the lattice is simultaneously displayed in a sum over lattice points with a dual link associated to that lattice point, whereas \eq{supmeas2} refers to only a single point on the moduli space. Note that the difference operator in the last term vanishes identically since there are no points on the lattice beyond $N^i=k$. The {\it  lattice measure} is defined by multiplying these factors for each PCO:
\begin{eqnarray}
{\bf X}_{\alpha_0...\alpha_k}^p \lineup\equiv \bigg[\Big(X(y^1_{\alpha_0})-(d+\Delta)\xi(y^1_{\alpha_0})\Big)\sum_{\vec{N},N^1=0}d(\vec{N}) + ...+\Big(X(y^1_{\alpha_k})-(d+\Delta)\xi(y^1_{\alpha_k})\Big)\sum_{\vec{N},N^1=k}d(\vec{N})\bigg]\times \nonumber\\
\lineup\ \ \  \ \ \ \ \ \ \ \ \ \ \ \ \ \ \ \ \ \ \ \ \ \ \ \ \ \ \ \ \ \ \ \ \ \ \ \ \ \ \vdots \ \ \ \nonumber\\
\lineup\ \ \ \times\bigg[\Big(X(y^p_{\alpha_0})-(d+\Delta)\xi(y^p_{\alpha_0})\Big)\sum_{\vec{N},N^p=0}d(\vec{N}) + ...+\Big(X(y^p_{\alpha_k})-(d+\Delta)\xi(y^p_{\alpha_k})\Big)\sum_{\vec{N},N^p=k}d(\vec{N}) \bigg].
\end{eqnarray}
This is a difference form of inhomogeneous degree, and is Grassmann even. We have the important property
\begin{equation}
(Q-d){\bf X}_{\alpha_0...\alpha_k}^p = \Delta {\bf X}_{\alpha_0...\alpha_k}^p.\phantom{\Bigg[}
\end{equation}
This follows since each factor $[Q-(d+\Delta)]$-exact in the large Hilbert space. For example, the $p$th factor can be written
\begin{eqnarray}
\lineup \bigg[\Big(X(y^p_{\alpha_0})-(d+\Delta)\xi(y^p_{\alpha_0})\Big)\sum_{\vec{N},N^p=0}d(\vec{N}) + ...+\Big(X(y^p_{\alpha_k})-(d+\Delta)\xi(y^p_{\alpha_k})\Big)\sum_{\vec{N},N^p=k}d(\vec{N})\bigg] \nonumber\\
\lineup\ \ \ \ \ \ \ \ \ \ \ \ \ \ \ \ \ \ \ \ \ \ \ \ \ \ \ \ \ \ \ \ \ \ \ \ \ \ \ \ \ \ \ \ \ =(Q-(d+\Delta))\bigg[\xi(y^p_{\alpha_0})\sum_{\vec{N},N^p=0}d(\vec{N}) + ...+\xi(y^p_{\alpha_k})\sum_{\vec{N},N^p=k}d(\vec{N})\bigg].\label{eq:QdDmeas}
\end{eqnarray}
This fact will be important in understanding the relation to the algebraic formalism.

It is helpful to see the lattice measure expanded into components. We will do this for the case of two PCOs and a codimension 2 face of the dual triangulation $M_{\alpha\beta\gamma}$ with $M_\alpha<M_\beta<M_\gamma$. The lattice measure is ${\bf X}^2_{\alpha\beta\gamma}$, and is defined on the $2\times 2$ lattice shown in figure \ref{fig:links}. To keep equations shorter, we will drop $d\xi$ terms which are always subtracted from $X$---interpreted literally, this would mean that the sections are constant in the coordinate~$y$. The difference 0-form part of the measure consists of $9$ terms:
\begin{eqnarray}
\lineup X(y^1_\alpha)X(y^2_\alpha)d(0,0) + X(y^1_\beta)X(y^2_\alpha)d(1,0)+X(y^1_\gamma)X(y^2_\alpha)d(2,0)\nonumber\\
\lineup +X(y^1_\alpha)X(y^2_\beta)d(0,1) + X(y^1_\beta)X(y^2_\beta)d(1,1)+X(y^1_\gamma)X(y^2_\beta)d(2,1)\nonumber\\
\lineup +X(y^1_\alpha)X(y^2_\gamma)d(0,2) + X(y^1_\beta)X(y^2_\gamma)d(1,2)+X(y^1_\gamma)X(y^2_\gamma)d(2,0).\label{eq:0Xabg}
\end{eqnarray}
These correspond to each of the $9$ points on the lattice. The difference 1-form part of the measure consists of $12$ terms:
\begin{eqnarray}
\lineup \Big(\xi(y^1_\beta)-\xi(y^1_\alpha)\Big)X(y^2_\alpha)d^1(0,0)+\Big(\xi(y^1_\gamma)-\xi(y^1_\beta)\Big)X(y^2_\alpha)d^1(1,0)\nonumber\\
\lineup +\Big(\xi(y^1_\beta)-\xi(y^1_\alpha)\Big)X(y^2_\beta)d^1(0,1)+\Big(\xi(y^1_\gamma)-\xi(y^1_\beta)\Big)X(y^2_\beta)d^1(1,1)\nonumber\\
\lineup +\Big(\xi(y^1_\beta)-\xi(y^1_\alpha)\Big)X(y^2_\gamma)d^1(0,2)+\Big(\xi(y^1_\gamma)-\xi(y^1_\beta)\Big)X(y^2_\gamma)d^1(1,2)\nonumber\\
\lineup +X(y_\alpha^1)\Big(\xi(y^2_\beta)-\xi(y^2_\alpha)\Big)d^2(0,0)+X(y_\alpha^1)\Big(\xi(y^2_\gamma)-\xi(y^2_\beta)\Big)d^2(0,1)\nonumber\\
\lineup +X(y_\beta^1)\Big(\xi(y^2_\beta)-\xi(y^2_\alpha)\Big)d^2(1,0)+X(y_\beta^1)\Big(\xi(y^2_\gamma)-\xi(y^2_\beta)\Big)d^2(1,1)\nonumber\\
\lineup +X(y_\gamma^1)\Big(\xi(y^2_\beta)-\xi(y^2_\alpha)\Big)d^2(2,0)+X(y_\gamma^1)\Big(\xi(y^2_\gamma)-\xi(y^2_\beta)\Big)d^2(2,1).
\end{eqnarray}
These correspond to each of the $12$ line segments connecting neighboring lattice sites. Finally, the difference 2-form part of the measure consists of $4$ terms:
\begin{eqnarray}
\lineup -\Big(\xi(y_\beta^1)-\xi(y_\alpha^1)\Big)\Big(\xi(y_\beta^2)-\xi(y_\alpha^2)\Big)d^{12}(0,0)-\Big(\xi(y_\gamma^1)-\xi(y_\beta^1)\Big)\Big(\xi(y_\beta^2)-\xi(y_\alpha^2)\Big)d^{12}(1,0)\nonumber\\
\lineup -\Big(\xi(y_\beta^1)-\xi(y_\alpha^1)\Big)\Big(\xi(y_\gamma^2)-\xi(y_\beta^2)\Big)d^{12}(0,1)-\Big(\xi(y_\gamma^1)-\xi(y_\beta^1)\Big)\Big(\xi(y_\gamma^2)-\xi(y_\beta^2)\Big)d^{12}(1,1).\label{eq:2Xabg}
\end{eqnarray}
The overall sign appears from commuting dual 1-forms $d^1(N^1,N^2)$ through $\xi$. The $4$ terms correspond to the four square regions of the lattice contained in neighboring lattice sites. Note that the lattice measure is manifestly defined in the small Hilbert space, since it is expressed directly in terms of differences of $\xi$s.

\begin{figure}
\begin{center}
\resizebox{3.2in}{2.6in}{\includegraphics{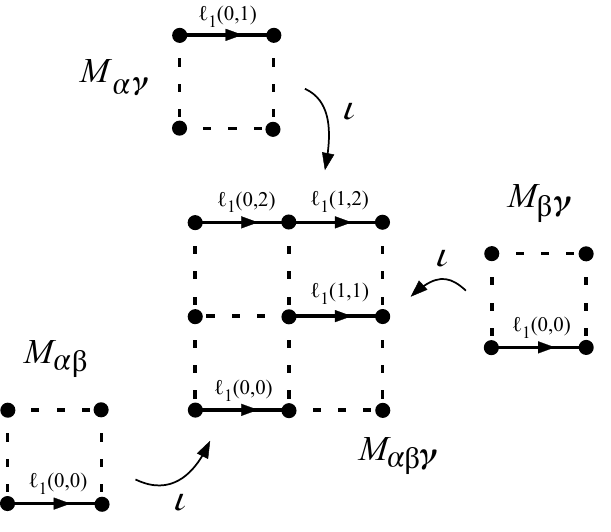}}
\end{center}
\caption{\label{fig:linkmap} The inclusion map sends links from the lattices of $M_{\alpha\beta},M_{\beta\gamma}$ and $M_{\alpha\gamma}$ into the lattice of $M_{\alpha\beta\gamma}$}
\end{figure}

In summary, for each face $M_{\alpha_0...\alpha_k}$ of the dual triangulation we have a cubic $p$-dimensional lattice with $(k+1)^p$ lattice sites, and a lattice measure ${\bf X}_{\alpha_0...\alpha_k}^p$ which acts on the links of that lattice. The Sen-Witten vertical corrections are given by acting the lattice measure on a lattice $k$-chain $C_{\alpha_0...\alpha_k}^p$:
\begin{equation}X_{\alpha_0...\alpha_k}^p = {\bf X}^p_{\alpha_0...\alpha_k}\Big[C^p_{\alpha_0...\alpha_k}\Big].\label{eq:SWX}\end{equation}
Note that the lattice chain $C_{\alpha_0...\alpha_k}^p$ must be built from $k$-dimensional links. Roughly speaking, this is because on a codimension $k$ face of the dual triangulation the integration cycle in $Y^p$ will contain $k$ dimensions tangent to the fiber. Gauge invariance imposes a condition on the choice of lattice chains. Recall from \eq{gaugeX} that the vertical corrections must satisfy 
\begin{equation}
(Q-d)X^p_{\alpha_0...\alpha_k} = (\delta X^p)_{\alpha_0...\alpha_k}.\label{eq:gaugeX2}
\end{equation}
Computing the left hand side gives
\begin{eqnarray}
(Q-d){\bf X}^p_{\alpha_0...\alpha_k}\Big[C^p_{\alpha_0...\alpha_k}\Big]\lineup =\Delta{\bf X}^p_{\alpha_0...\alpha_k}\Big[C^p_{\alpha_0...\alpha_k}\Big]\nonumber\\
\lineup = (-1)^k{\bf X}^p_{\alpha_0...\alpha_k}\Big[\d C^p_{\alpha_0...\alpha_k}\Big],
\end{eqnarray}
where the sign comes from the Grassmann parity of $C_{\alpha_0...\alpha_k}^p$. Expanding the right hand side of \eq{gaugeX2} we learn that
\begin{equation}
(-1)^k{\bf X}^p_{\alpha_0...\alpha_k}\Big[\d C^p_{\alpha_0...\alpha_k}\Big] = \sum_{n=0}^k (-1)^n{\bf X}^p_{\alpha_0...\widehat{\alpha}_n...\alpha_k}\Big[ C^p_{\alpha_0...\widehat{\alpha}_n...\alpha_k}\Big].\label{eq:Cconst}
\end{equation}
To understand the implications of this equation we need to compare chains defined on different lattices. This can be done in a natural way by identifying points between lattices which refer to the same polyhedra. As an example, let us consider a codimension 2 face $M_{\alpha\beta\gamma}$ which appears at the common boundary of codimension 1 faces $M_{\alpha\beta},M_{\beta\gamma}$ and $M_{\alpha\gamma}$. We assume that $M_\alpha<M_\beta<M_\gamma$. If we have two PCOs, $M_{\alpha\beta\gamma}$ comes with a $2\times 2$ lattice as shown in figure \ref{fig:links}, while $M_{\alpha\beta},M_{\beta\gamma}$ and $M_{\alpha\gamma}$ come with $1\times 1$ lattices. Now consider the following:
\begin{itemize}
\item The link $\ell_1(0,0)$ in the lattice of $M_{\alpha\beta}$ connects the points $(0,0)$ and $(1,0)$. These points can be equivalently labeled by pairs of polyhedra $(M_\alpha,M_\alpha)$ and $(M_\beta,M_\alpha)$, respectively. In the lattice of $M_{\alpha\beta\gamma}$, these pairs of polyhedra correspond to $(0,0)$ and $(0,1)$. Therefore the link $\ell_1(0,0)$ of the lattice of $M_{\alpha\beta}$ is naturally paired with the same link $\ell_1(0,0)$ of the lattice of $M_{\alpha\beta\gamma}$.
\item The link $\ell_1(0,0)$ in the lattice of $M_{\beta\gamma}$ connects the points $(0,0)$ and $(1,0)$. These points can be equivalently labeled by pairs of polyhedra $(M_\beta,M_\beta)$ and $(M_\gamma,M_\beta)$, respectively. In the lattice of $M_{\alpha\beta\gamma}$, these pairs of polyhedra correspond to $(1,1)$ and $(2,1)$. Therefore the link $\ell_1(0,0)$ of the lattice of $M_{\beta\gamma}$ is naturally paired with the link $\ell_1(1,1)$ of the lattice of $M_{\alpha\beta\gamma}$.
\item The link $\ell_1(0,1)$ in the lattice of $M_{\alpha\gamma}$ connects the points $(0,1)$ and $(1,1)$. These points can be equivalently labeled by pairs of polyhedra $(M_\alpha,M_\gamma)$ and $(M_\gamma,M_\gamma)$, respectively. In the lattice of $M_{\alpha\beta\gamma}$, these pairs of polyhedra correspond to $(0,2)$ and $(2,2)$. These two points are not neighboring lattice sites, and there is no link connecting them. However, we may connect these points by a sum of links $\ell_1(0,2)+\ell_1(1,2)$. Therefore the link $\ell_1(0,1)$ in the lattice of $M_{\alpha\gamma}$ is naturally paired with a 1-chain $\ell_1(0,2)+\ell_1(1,2)$ on the lattice of~$M_{\alpha\beta\gamma}$.
\end{itemize}
This is shown in figure  \ref{fig:linkmap}. The important property of these identifications is that the lattice measures evaluated on the respective links are equal: 
\begin{eqnarray}
{\bf X}^2_{\alpha\beta}[\ell_1(0,0)] \lineup = {\bf X}^2_{\alpha\beta\gamma}[\ell_1(0,0)] \nonumber\\
{\bf X}^2_{\beta\gamma}[\ell_1(0,0)] \lineup = {\bf X}^2_{\alpha\beta\gamma}[\ell_1(1,1)] \nonumber\\
{\bf X}^2_{\alpha\gamma}[\ell_1(0,1)] \lineup = {\bf X}^2_{\alpha\beta\gamma}[\ell_1(0,2)+\ell_1(1,2)],
\end{eqnarray}
where both sides of the equations are evaluated at a common point $m\in M_{\alpha\beta\gamma}$. Now let us describe how this works in general. Suppose that a codimension $l$ face $M_{\beta_0...\beta_l}$ has a boundary which intersects a codimension $k$ face $M_{\alpha_0...\alpha_k}$ with $k>l$. This means that there is an inclusion map $\iota$ from the integers $0,...,l$ to the integers $0,...,k$ satisfying
\begin{equation}
M_{\beta_n} = M_{\alpha_{\iota(n)}}
\end{equation}
and
\begin{equation}
\iota(n_1)<\iota(n_2)\ \ \mathrm{iff}\ \ n_1<n_2.
\end{equation}
The link $\ell_{i_1...i_h}(\vec{\widetilde{N}})$ on the lattice of $M_{\beta_0...\beta_l}$ maps to a sum of links $\ell_{i_1...i_h}(\vec{N})$ on the lattice of $M_{\alpha_0...\alpha_k}$ with
\begin{eqnarray}
\lineup \iota(\widetilde{N}^{i_1})\leq N^{i_1}< \iota(\widetilde{N}^{i_1}+1)\nonumber\\
\lineup\ \ \ \ \ \ \ \ \ \ \ \ \ \ \vdots\nonumber\\
\lineup \iota(\widetilde{N}^{i_h})\leq N^{i_h}< \iota(\widetilde{N}^{i_h}+1)\nonumber\\
\lineup N^i = \iota(\widetilde{N}^i),\ \ \ \ i\neq i_1,...,i_h.
\end{eqnarray}
We will denote the resulting sum of links as $\iota\circ\ell_{i_1...i_h}(\vec{\widetilde{N}})$. As in the above examples, the inclusion map $\iota$ has the property that the lattice measures acting on the respective lattice chains are equal: 
\begin{equation}{\bf X}^p_{\beta_0...\beta_l}[C] = {\bf X}^p_{\alpha_0...\alpha_k}[\iota\circ C],\label{eq:sigmameas}\end{equation}
where both sides are evaluated at a common point $m\in M_{\alpha_0...\alpha_k}$. Using this map, we can express all terms in \eq{Cconst} using the common lattice measure acting on chains in the same lattice:
\begin{equation}
(-1)^k{\bf X}^p_{\alpha_0...\alpha_k}\Big[\d C^p_{\alpha_0...\alpha_k}\Big] = \sum_{n=0}^k (-1)^n{\bf X}^p_{\alpha_0...\alpha_k}\Big[\iota\circ C^p_{\alpha_0...\widehat{\alpha}_n...\alpha_k}\Big].
\end{equation}
This implies
\begin{equation}\d C^p_{\alpha_0...\alpha_k} = \sum_{n=0}^k(-1)^{n+k}\iota\circ C^p_{\alpha_0...\widehat{\alpha}_n...\alpha_k}.\label{eq:Cconst2}\end{equation}
Therefore the boundary of the lattice chain on a codimension $k$ face must be given by patching together the lattice chains on faces of one lower codimension. In principle, there could be a topological obstruction to solution of this equation from the homology of $\d$. Since the lattices we consider do not have ``holes," the only possible homology group appears for $0$-chains. The lattice $0$-chains are associated with the codimension $0$ polyhedra of the dual triangulation, each of which carries a lattice consisting of only one point. Each of these lattices carries a single $0$-link,
\begin{equation}\ell(0,...,0),\end{equation}
and the most general choice of lattice $0$-chain would multiply this link by an integer which could in principle be different on each face of the dual triangulation. However, if the 0-chain is chosen differently on any two polyhedra, there will be an obstruction to the solution of \eq{Cconst2} from homology. Therefore the $0$-chains of all polyhedra must be equal, and with an appropriate normalization of the amplitude they can be set equal to $\ell(0,...,0)$. What we have just argued is that the measure on each codimension $0$ face $M_\alpha$ must be given the pullback of the superstring measure,
\begin{equation}\Big[X(y^1_\alpha) - d\xi(y^1_\alpha)\Big]\ .\,.\,.\ \Big[X(y^p_\alpha) - d\xi(y^p_\alpha)\Big],\end{equation}
and not the measure multiplied by an integer which may take different values between different polyhedra. Of course, this is what we have been assuming from the beginning.

\subsection{Relation to the Algebraic Approach}

To relate the Sen-Witten construction to the algebraic approach, we must derive a gauge amplitude $\langle \alpha^p|$ and an amplitude $\langle A^{p-1}|$ with one fewer PCO insertion satisfying
\begin{eqnarray}
\langle A^p| \lineup = \langle\alpha^p|Q\nonumber\\
\langle \alpha^p|\eta \lineup = \langle A^{p-1}|,
\end{eqnarray}
where the original amplitude $\langle A^p|$ is defined with Sen-Witten vertical corrections. Assuming that $\langle A^{p-1}|$ is also characterized by Sen-Witten vertical corrections, we may continue in the same way and derive a hierarchy of amplitudes and gauge amplitudes down to $\langle A^0|$, where PCOs are absent. For definiteness, we will assume that the $i$th amplitude in this hierarchy contains $i$ PCOs with positions $y_\alpha^1,...,y_\alpha^i$. Thus, the amplitude $\langle A^1|$ contains one PCO with position $y_\alpha^1$, $\langle A^2|$ contains two PCOs with positions $y_\alpha^1,y_\alpha^2$, and so on until $\langle A^p|$ contains all PCOs with positions $y_\alpha^1,...,y_\alpha^p$.

To derive the gauge amplitude $\langle \alpha^p|$, we observe that the lattice measure can be expressed in the form
\begin{equation}{\bf X}^p_{\alpha_0...\alpha_k} = (Q-(d+\Delta))\bm{\Xi}_{\alpha_0...\alpha_k}^p.\label{eq:XXi}\end{equation}
The object $\bm{\Xi}_{\alpha_0...\alpha_k}^p$ on the right hand side will be called the {\it gauge lattice measure}. It takes the same form as ${\bf X}_{\alpha_0...\alpha_k}^p$, except that the $p$th factor is replaced using \eq{QdDmeas}:
\begin{eqnarray}
\bm{\Xi}_{\alpha_0...\alpha_k}^p \lineup\equiv \bigg[\Big(X(y^1_{\alpha_0})-(d+\Delta)\xi(y^1_{\alpha_0})\Big)\sum_{\vec{N},N^1=0}d(\vec{N}) + ...+\Big(X(y^1_{\alpha_k})-(d+\Delta)\xi(y^1_{\alpha_k})\Big)\sum_{\vec{N},N^1=k}d(\vec{N})\bigg]\times \nonumber\\
\lineup\ \ \  \ \ \ \ \ \ \ \ \ \ \ \ \ \ \ \ \ \ \ \ \ \ \ \ \ \ \ \ \ \ \ \ \ \ \ \ \ \ \vdots \ \ \ \nonumber\\
\lineup\ \ \ \times\bigg[\Big(X(y^{p-1}_{\alpha_0})-(d+\Delta)\xi(y^{p-1}_{\alpha_0})\Big)\sum_{\vec{N},N^{p-1}=0}d(\vec{N}) + ...+\Big(X(y^{p-1}_{\alpha_k})-(d+\Delta)\xi(y^{p-1}_{\alpha_k})\Big)\sum_{\vec{N},N^{p-1}=k}d(\vec{N}) \bigg]\nonumber\\
\lineup \ \ \ \times\bigg[\xi(y^p_{\alpha_0})\sum_{\vec{N},N^p=0}d(\vec{N}) + ...+\xi(y^p_{\alpha_k})\sum_{\vec{N},N^p=k}d(\vec{N})\bigg].\label{eq:Xi}
\end{eqnarray}
This is a difference form of inhomogeneous degree, is Grassmann odd, and is defined in the large Hilbert space. The fact that the $p$th factor plays a special role is related to our assumption that the amplitude $\langle A^{p-1}|$ should not contain the PCO with coordinate $y_\alpha^p$. Let us expand the gauge lattice measure out for the case of 2 PCOs and a codimension 2 face $M_{\alpha\beta\gamma}$ with $M_\alpha<M_\beta<M_\gamma$. As before, we will suppress $d\xi$ terms which are always subtracted from $X$. The difference $0$-form part of ${\bf \Xi}^2_{\alpha\beta\gamma}$ consists of $9$ terms:
\begin{eqnarray}
\lineup X(y^1_\alpha)\xi(y^2_\alpha)d(0,0) + X(y^1_\beta)\xi(y^2_\alpha)d(1,0)+X(y^1_\gamma)\xi(y^2_\alpha)d(2,0)\nonumber\\
\lineup +X(y^1_\alpha)\xi(y^2_\beta)d(0,1) + X(y^1_\beta)\xi(y^2_\beta)d(1,1)+X(y^1_\gamma)\xi(y^2_\beta)d(2,1)\nonumber\\
\lineup +X(y^1_\alpha)\xi(y^2_\gamma)d(0,2) + X(y^1_\beta)\xi(y^2_\gamma)d(1,2)+X(y^1_\gamma)\xi(y^2_\gamma)d(2,0),\label{eq:0Xiabg}
\end{eqnarray}
and the difference $1$-form part consists of $6$ terms:
\begin{eqnarray}
\lineup -\Big(\xi(y^1_\beta)-\xi(y^1_\alpha)\Big)\xi(y^2_\alpha)d^1(0,0)-\Big(\xi(y^1_\gamma)-\xi(y^1_\beta)\Big)\xi(y^2_\alpha)d^1(1,0)\nonumber\\
\lineup -\Big(\xi(y^1_\beta)-\xi(y^1_\alpha)\Big)\xi(y^2_\beta)d^1(0,1)-\Big(\xi(y^1_\gamma)-\xi(y^1_\beta)\Big)\xi(y^2_\beta)d^1(1,1)\nonumber\\
\lineup -\Big(\xi(y^1_\beta)-\xi(y^1_\alpha)\Big)\xi(y^2_\gamma)d^1(0,2)-\Big(\xi(y^1_\gamma)-\xi(y^1_\beta)\Big)\xi(y^2_\gamma)d^1(1,2).\label{eq:1Xiabg}
\end{eqnarray}
There is no difference $2$-form component. Note that the dual links $d^2(N^1,N^2)$ are absent. This means that the action of $\bm{\Xi}^2_{\alpha\beta\gamma}$ on lattice chains will be independent of the links parallel to the axis of the second PCO. 

Using the gauge lattice measure we can reexpress the Sen-Witten vertical correction as follows: 
\begin{eqnarray}
X_{\alpha_0...\alpha_k}^p \lineup = (Q-d-\Delta)\bm{\Xi}^p_{\alpha_0...\alpha_k}[C_{\alpha_0...\alpha_k}^p]\nonumber\\
\lineup = (Q-d)\bm{\Xi}^p_{\alpha_0...\alpha_k}[C_{\alpha_0...\alpha_k}^p]+(-1)^k \bm{\Xi}^p_{\alpha_0...\alpha_k}[\d C_{\alpha_0...\alpha_k}^p]\nonumber\\
\lineup = (Q-d)\bm{\Xi}^p_{\alpha_0...\alpha_k}[C_{\alpha_0...\alpha_k}^p]+\sum_{n=0}^k (-1)^n \bm{\Xi}^p_{\alpha_0...\alpha_k}[\iota\circ C_{\alpha_0...\widehat{\alpha}_n...\alpha_k}^p]\nonumber\\
\lineup = (Q-d)\bm{\Xi}^p_{\alpha_0...\alpha_k}[C_{\alpha_0...\alpha_k}^p]+\sum_{n=0}^k (-1)^n \bm{\Xi}^p_{\alpha_0...\widehat{\alpha}_n...\alpha_k}[C_{\alpha_0...\widehat{\alpha}_n...\alpha_k}^p].\label{eq:SWgaugexi}
\end{eqnarray}
In the last step we noted that the analogue of \eq{sigmameas} also holds for the gauge lattice measure:
\begin{equation}\bm{\Xi}_{\beta_0...\beta_l}^p[C] = \bm{\Xi}_{\alpha_0...\alpha_k}^p[\iota\circ C],\end{equation}
where $M_{\beta_0}<...<M_{\beta_l}$ are a subset of the polyhedra $M_{\alpha_0}<...<M_{\alpha_k}$. Comparing \eq{SWgaugexi} to \eq{gaugexi}, we can identify the operator insertions $\Xi_{\alpha_0...\alpha_k}^p$ defining the gauge amplitude:
\begin{equation}
\Xi_{\alpha_0...\alpha_k}^p = \bm{\Xi}_{\alpha_0...\alpha_k}^p[C_{\alpha_0...\alpha_k}^p].\label{eq:SWXi}
\end{equation}
Thus we have a gauge amplitude $\langle \alpha^p|$ whose BRST variation gives an amplitude with Sen-Witten vertical corrections. 

\begin{figure}
\begin{center}
\resizebox{5.5in}{1.1in}{\includegraphics{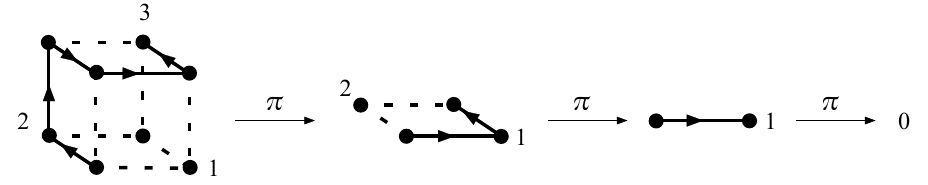}}
\end{center}
\caption{\label{fig:projection} An amplitude containing $p$ PCOs defined with Sen-Witten vertical corrections can be constructed algebraically from a sequence of amplitudes containing $0,1,2,...,p-1$ PCOs. The amplitudes with $0,...,p-1$ PCOs can also be characterized by Sen-Witten vertical corrections; the appropriate lattice chains can be obtained by projection of the chains of the original amplitude down to lower dimensional lattices. Shown above is an example of chains for amplitudes with $3,2,1$ and $0$ PCOs on a codimension $1$ face of the dual triangulation of moduli space. The projections sequentially eliminate the $3$-, $2$-, and finally $1$-axis, which maps the 1-chain defining the vertical correction to zero. Consider the projection from the 3 PCO lattice down to the $2$ PCO lattice. In this case we are projecting out the vertical direction corresponding to the third PCO. On the leftmost face of the cube, the link $\ell_3(0,1,0)$ is annihilated by the projection, and the links $\ell_2(0,0,0)$ and $-\ell_2(0,0,1)$ cancel after the projection. What is left is the chain $\ell_1(0,0,1)+\ell_2(1,0,1)$, which maps to $\ell_1(0,0)+\ell_2(1,0)$ on the $2$ PCO lattice.}
\end{figure}

The next step is to derive the amplitude $\langle A^{p-1}|$ with one fewer PCO. If the vertical corrections of $\langle A^{p-1}|$ take the Sen-Witten form, we must have
\begin{equation} \eta\bm{\Xi}^p_{\alpha_0...\alpha_k}[C^p_{\alpha_0...\alpha_k}] = {\bf X}^{p-1}_{\alpha_0...\alpha_k}[C^{p-1}_{\alpha_0...\alpha_k}],\end{equation}
for some $k$-chain $C^{p-1}_{\alpha_0...\alpha_k}$ on a $(p-1)$-dimensional lattice. If we compute $\eta\bm{\Xi}_{\alpha_0...\alpha_k}^p$, the $p$th factor in \eq{Xi} is replaced with the identity, and the dependence on the coordinate $y_\alpha^p$ drops out, as we anticipated. What is left is almost the same as the lattice measure ${\bf X}^{p-1}_{\alpha_0...\alpha_k}$. However, $\eta\bm{\Xi}^p_{\alpha_0...\alpha_k}$ is defined on a lattice with an additional coordinate $N^p$. One can check that $\eta\bm{\Xi}^p_{\alpha_0...\alpha_k}$ is related to ${\bf X}^{p-1}_{\alpha_0...\alpha_k}$ by replacing the dual links on the $(p-1)$-dimensional lattice according to 
\begin{equation}d^{i_1...i_h}(N^1,...,N^{p-1})\ \ \to\ \ \sum_{N^p=0}^k d^{i_1...i_h}(N^1,...,N^{p-1},N^p),\end{equation}
where all $i_1,...,i_h\neq p$. The net effect of this replacement is to define a projection map from lattice chains on the $p$-dimensional lattice down to lattice chains on the $(p-1)$-dimensional lattice given by
\begin{eqnarray}
\pi\circ \ell_{i_1...i_h}(N^1,...,N^{p-1},N^p) \lineup = \ell_{i_1...i_h}(N^1,...,N^{p-1}) \ \ \ \ \ \mathrm{if\ all}\ i_1,...,i_h\neq p\nonumber\\
\pi\circ \ell_{i_1...i_h}(N^1,...,N^{p-1},N^p) \lineup = 0,\ \ \ \ \ \ \ \ \ \ \ \ \ \ \ \ \ \ \ \ \ \ \ \ \ \ \ \ \ \mathrm{if\ any}\ i_1,...,i_h\ \mathrm{is\ equal\ to}\ p.
\end{eqnarray}
The projection effectively forgets about links parallel to the $p$ direction, and links orthogonal to the $p$ direction sharing a common $p$th coordinate are superimposed. An example is shown in figure \ref{fig:projection}. With this projection map, we have the equality
\begin{equation}\eta\bm{\Xi}^p_{\alpha_0...\alpha_k}[C^p_{\alpha_0...\alpha_k}] = {\bf X}^{p-1}_{\alpha_0...\alpha_k}[\pi\circ C^p_{\alpha_0...\alpha_k}].\end{equation}
Therefore, the lattice chains defining the vertical corrections of $\langle A^{p-1}|$ are given by
\begin{equation}C^{p-1}_{\alpha_0...\alpha_k} = \pi\circ C^p_{\alpha_0...\alpha_k}.\end{equation}
One can check that the projection map commutes through the boundary operator and inclusion map, so $C^{p-1}_{\alpha_0...\alpha_k}$ satisfies \eq{Cconst2} if $C^p_{\alpha_0...\alpha_k}$ does. Since $\langle A^{p-1}|$ is defined by Sen-Witten vertical corrections, we may continue following the above discussion to derive the full hierarchy of amplitudes and gauge amplitudes. The amplitudes are simply given by appropriately projecting the chains $C^p_{\alpha_0...\alpha_k}$ down to lower dimensional lattices. See figure \ref{fig:projection}.

This explains how the algebraic construction of the amplitude may be rederived once the Sen-Witten construction of the amplitude has been provided. However, suppose we want to go in reverse, using the algebraic construction to derive an amplitude with Sen-Witten vertical corrections. Suppose we have succeeded in doing this up to the amplitude $\langle A^{p-1}|$ with $p-1$ PCO insertions. This amplitude is then characterized by a collection of lattice chains $C^{p-1}_{\alpha_0...\alpha_k}$ satisfying \eq{Cconst2} for each face of the dual triangulation. To derive the amplitude $\langle A^p|$, we must find a collection of lattice chains $\widetilde{C}^p_{\alpha_0...\alpha_k}$ in $p$-dimensional lattices satisfying
\begin{equation} \eta\bm{\Xi}^p_{\alpha_0...\alpha_k}[\widetilde{C}^p_{\alpha_0...\alpha_k}] = {\bf X}^{p-1}_{\alpha_0...\alpha_k}[C^{p-1}_{\alpha_0...\alpha_k}].\label{eq:CtC}\end{equation}
The operator insertions defining the gauge amplitude are given by
\begin{equation}\Xi_{\alpha_0...\alpha_k}^p = \bm{\Xi}_{\alpha_0...\alpha_k}^p[\widetilde{C}_{\alpha_0...\alpha_k}^p].\end{equation}
We may construct the amplitude $\langle A^p|$ from the vertical corrections
\begin{equation}X_{\alpha_0...\alpha_k}^p =(Q-d)\Xi_{\alpha_0...\alpha_k}^p +(\delta \Xi^p)_{\alpha_0...\alpha_k}.\label{eq:gaugexi2}\end{equation}
This however raises a puzzle. The only condition that \eq{CtC} imposes on the lattice chains $\widetilde{C}^p_{\alpha_0...\alpha_k}$ is that they project down to the lattice chains of $\langle A^{p-1}|$:
\begin{equation}\pi\circ \widetilde{C}^p_{\alpha_0...\alpha_k} = C^{p-1}_{\alpha_0...\alpha_k} .\end{equation}
However, the chains $\widetilde{C}^p_{\alpha_0...\alpha_k}$ do not necessarily satisfy \eq{Cconst2}. This means that vertical corrections given~by
\begin{equation}{\bf X}^p_{\alpha_0...\alpha_k}[\widetilde{C}^p_{\alpha_0...\alpha_k}]\end{equation}
do not define a gauge invariant amplitude. The resolution to this puzzle is that \eq{gaugexi2} does not produce vertical corrections characterized by the lattice chains $\widetilde{C}^p_{\alpha_0...\alpha_k}$. Rather, they produce vertical corrections corresponding to a different set of lattice chains $C^p_{\alpha_0...\alpha_k}$ which sew together consistently with \eq{Cconst2} and satisfy
\begin{equation} \bm{\Xi}^p_{\alpha_0...\alpha_k}[C^p_{\alpha_0...\alpha_k}] =  \bm{\Xi}^p_{\alpha_0...\alpha_k}[\widetilde{C}^p_{\alpha_0...\alpha_k}].\end{equation}
This is possible since the gauge lattice measure $\bm{\Xi}^p_{\alpha_0...\alpha_k}$ is degenerate: it vanishes acting on links with components parallel to the $p$-axis. So effectively \eq{gaugexi2}  ``fills in" the missing links of $\widetilde{C}^p_{\alpha_0...\alpha_k}$ to define new chains $C^p_{\alpha_0...\alpha_k}$ consistent with gauge invariance. 

\subsection{Examples}

Let us give examples to illustrate the discussion of the last subsection. We start by constructing an amplitude with a single PCO. The structure is somewhat degenerate in this example, and the result is the same as in subsection~\ref{subsec:examples}, but it is useful to see how it works.  For an amplitude containing no PCOs, we can define Sen-Witten ``vertical corrections" using a $0$-dimensional lattice. On a $0$-dimensional lattice, there is a $0$-link $\ell$ and a dual $0$-link $d$ satisfying 
\begin{equation}
d[\ell]=1.
\end{equation}
The link and dual link do not carry any lattice coordinates. Since the amplitude carries no PCOs, the lattice measure on each face of the dual triangulation is given by
\begin{equation}
{\bf X}^0_{\alpha_0...\alpha_k} = d.
\end{equation}
The lattice chains which define the vertical corrections take the form:
\begin{eqnarray}
C_\alpha^0 \lineup = \ell\nonumber\\
C_{\alpha\beta}^0 \lineup = 0\nonumber\\
C^0_{\alpha\beta\gamma}\lineup = 0\nonumber\\
\lineup \vdots\ \ .
\end{eqnarray}
The chains on higher codimension faces vanish because the $0$ dimensional lattice does not support higher dimensional links. Therefore, the ``vertical corrections" of an amplitude without PCOs are given by
\begin{eqnarray}
X^0_\alpha\lineup =\ \ {\bf X}^0_{\alpha}[C_\alpha^0]  \ \ \ \,= 1\nonumber\\
X^0_{\alpha\beta}\lineup = \ {\bf X}^0_{\alpha\beta}[C_{\alpha\beta}^0] \ \, = 0\nonumber\\
X^0_{\alpha\beta\gamma}\lineup = {\bf X}^0_{\alpha\beta\gamma}[C_{\alpha\beta\gamma}^0] = 0\nonumber\\
\lineup\ \ \ \ \ \ \ \ \ \ \vdots\ \ .
\end{eqnarray}
This agrees with \eq{X0}. To construct the amplitude with a single PCO, we introduce 1-dimensional lattices with lattice sites $0,...,k$ for each codimension $k$ face of the dual triangulation. The vertical corrections are defined by lattice chains $C_{\alpha_0...\alpha_k}^1$ satisfying
\begin{equation}\pi\circ C_{\alpha_0...\alpha_k}^1 = C_{\alpha_0...\alpha_k}^0.\label{eq:piC1C0}\end{equation}
We can choose a solution of the form
\begin{eqnarray}
C^1_\alpha\lineup = \ell(0)\nonumber\\
C^1_{\alpha\beta}\lineup = \ell_1(0)\nonumber\\
C^1_{\alpha\beta\gamma}\lineup = 0\nonumber\\
\lineup \vdots\ \ .\label{eq:C1ex}
\end{eqnarray}
The chains on faces of codimension $2$ and higher vanish because the $1$-dimensional lattice does not support links of dimension $2$ and higher. The solution for $C^1_\alpha$ is uniquely determined by \eq{piC1C0}, but there is some ambiguity in the choice of $C^1_{\alpha\beta}$---any 1-chain will map to zero after the projection. We have chosen $C^1_{\alpha\beta}$ so that \eq{Cconst2} is satisfied:
\begin{equation}\d C^1_{\alpha\beta} = -\iota\circ C^1_\beta+\iota\circ C^1_\alpha.\label{eq:C1const}\end{equation}
The lattice measure and gauge lattice measure take the form: 
\begin{eqnarray}
{\bf X}_{\alpha_0...\alpha_k}^1 \lineup = \Big[X(y_{\alpha_0}^1)-d\xi(y_{\alpha_0}^1)\Big]d(0)+...+\Big[X(y_{\alpha_k}^1)-d\xi(y_{\alpha_k}^1)\Big]d(k)\nonumber\\
\lineup\ \ \ +\Big[\xi(y_{\alpha_1}^1)-\xi(y_{\alpha_0}^1)\Big]d^1(0)+...+\Big[\xi(y_{\alpha_k}^1)-\xi(y_{\alpha_{k-1}}^1)\Big]d^1(k-1)\nonumber\\
\bm{\Xi}_{\alpha_0...\alpha_k}^1\lineup = \xi(y_{\alpha_0}^1)d(0)+...+\xi(y_{\alpha_k}^1)d(k),
\end{eqnarray}
where $y^1_\alpha$ gives the location of the PCO as a function of $m\in M_\alpha$. The operator insertions defining the gauge amplitude are given by
\begin{eqnarray}
\Xi^1_\alpha\lineup =\ \ \bm{\Xi}^1_{\alpha}[C_\alpha^1]  \ \ \ \,= \xi(y^1_\alpha) \nonumber\\
\Xi^1_{\alpha\beta}\lineup = \ \bm{\Xi}^1_{\alpha\beta}[C_{\alpha\beta}^1] \ \, = 0\nonumber\\
\Xi^1_{\alpha\beta\gamma}\lineup = \bm{\Xi}^1_{\alpha\beta\gamma}[C_{\alpha\beta\gamma}^1] = 0.\nonumber\\
\lineup\ \ \ \ \ \ \ \ \ \ \vdots
\end{eqnarray}
This agrees with \eq{xi1}. We can derive the vertical corrections by substituting into
\begin{equation}X_{\alpha_0...\alpha_k}^1 =(Q-d)\Xi_{\alpha_0...\alpha_k}^1 +(\delta \Xi^1)_{\alpha_0...\alpha_k}.\end{equation}
However, since we have chosen $C^1_{\alpha\beta}$ to satisfy \eq{C1const}, we know that the answer will simply be given by acting the lattice measure on the respective lattice chains: 
\begin{eqnarray}
X^1_\alpha\lineup =\ \ {\bf X}^1_{\alpha}[C_\alpha^1]  \ \ \ \,= X(y_{\alpha}^1)-d\xi(y_\alpha^1)\nonumber\\
X^1_{\alpha\beta}\lineup = \ {\bf X}^1_{\alpha\beta}[C_{\alpha\beta}^1] \ \, = \xi(y_\beta^1)-\xi(y_\alpha^1)\nonumber\\
X^1_{\alpha\beta\gamma}\lineup = {\bf X}^1_{\alpha\beta\gamma}[C_{\alpha\beta\gamma}^1] = 0\nonumber\\
\lineup\ \ \ \ \ \ \ \ \ \ \vdots\ \ .
\end{eqnarray}
This agrees with \eq{X1ab}.

\begin{figure}
\begin{center}
\resizebox{3in}{2.5in}{\includegraphics{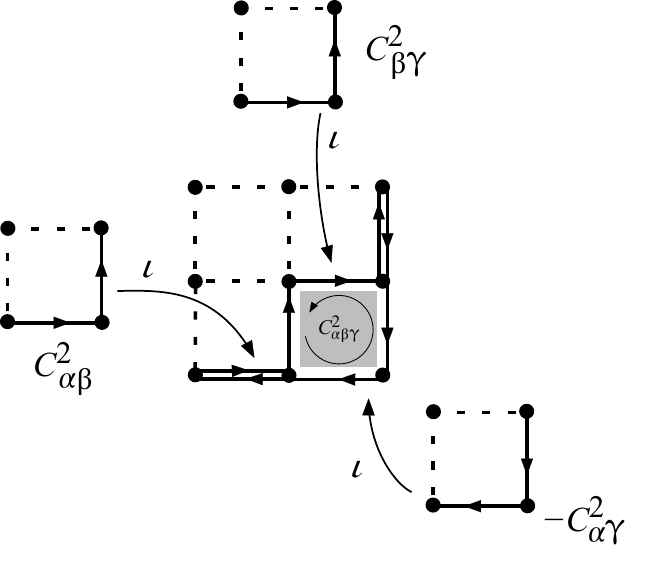}}
\end{center}
\caption{\label{fig:ex} The lattice chains in \eq{Cconstex} sew together consistently with gauge invariance.}
\end{figure}

Now let's consider the case of two PCOs. We must find chains on two dimensional lattices which satisfy
\begin{equation}\pi\circ\widetilde{C}^2_{\alpha_0...\alpha_k}=C^1_{\alpha_0...\alpha_k}.\label{eq:CtC2}\end{equation}
This time we will not attempt to choose lattice chains which solve the constraint \eq{Cconst2}. We will let the algebraic construction solve the constraint for us. A simple solution to \eq{CtC2} takes the form
\begin{eqnarray}
\widetilde{C}^2_\alpha\lineup = \ell(0,0)\nonumber\\
\widetilde{C}^2_{\alpha\beta}\lineup = \ell_1(0,0)\nonumber\\
\widetilde{C}^2_{\alpha\beta\gamma}\lineup = 0\nonumber\\
\widetilde{C}^2_{\alpha\beta\gamma\delta}\lineup = 0\nonumber\\\label{eq:C2t}
\lineup\vdots\ \ .
\end{eqnarray}
This is identical to \eq{C1ex} if we forget about the second lattice coordinate. The chains on faces of codimension $3$ and higher must vanish since a $2$-dimensional lattice does not support chains of dimension $3$ and higher. The choice of $0$-chains $\widetilde{C}^2_\alpha$ is unique, but there is ambiguity in the choice of $1$- and $2$-chains. Note that we have chosen the $1$- and $2$-chains in a form which is independent of the face of the dual triangulation. This is not necessary, but simplifies the computation. The lattice measure on faces of codimension $0$ and $1$ take the form:
\begin{eqnarray}
{\bf X}_\alpha^2\lineup = \Big[X(y_\alpha^1)-d\xi(y_\alpha^1)\Big]\Big[X(y_\alpha^2)-d\xi(y_\alpha^2)\Big]d(0,0)\nonumber\\
{\bf X}_{\alpha\beta}^2\lineup = \Big[X(y_\alpha^1)-d\xi(y_\alpha^1)\Big]\Big[X(y_\alpha^2)-d\xi(y_\alpha^2)\Big]d(0,0)+\Big[X(y_\beta^1)-d\xi(y_\beta^1)\Big]\Big[X(y_\alpha^2)-d\xi(y_\alpha^2)\Big]d(1,0)\nonumber\\
\lineup \ \ \ +\Big[X(y_\alpha^1)-d\xi(y_\alpha^1)\Big]\Big[X(y_\beta^2)-d\xi(y_\beta^2)\Big]d(0,1)+\Big[X(y_\beta^1)-d\xi(y_\beta^1)\Big]\Big[X(y_\beta^2)-d\xi(y_\beta^2)\Big]d(1,1)\nonumber\\
\lineup\ \ \ +\Big[\xi(y_\beta^1)-\xi(y_\alpha^1)\Big]\Big[X(y_\alpha^2)-d\xi(y_\alpha^2)\Big]d^1(0,0)+\Big[X(y_\alpha^1)-d\xi(y_\alpha^1)\Big]\Big[\xi(y_\beta^2)-\xi(y_\alpha^2)\Big]d^2(0,0)\nonumber\\
\lineup\ \ \ +\Big[X(y_\beta^1)-d\xi(y_\beta^1)\Big]\Big[\xi(y_\beta^2)-\xi(y_\alpha^2)\Big]d^2(1,0)+\Big[\xi(y_\beta^1)-\xi(y_\alpha^1)\Big]\Big[X(y_\beta^2)-d\xi(y_\beta^2)\Big]d^1(0,1)\nonumber\\
\lineup\ \ \ -\Big[\xi(y_\beta^1)-\xi(y_\alpha^1)\Big]\Big[\xi(y_\beta^2)-\xi(y_\alpha^2)\Big]d^{12}(0,0).
\end{eqnarray}
The lattice measure on faces of codimension $2$ is given in \eq{0Xabg}-\eq{2Xabg}, after subtracting $d\xi$ from $X$. The gauge lattice measure on faces of codimension $0$ and $1$ take the form
\begin{eqnarray}
\bm{\Xi}^2_\alpha\lineup = \Big[X(y_\alpha^1)-d\xi(y_\alpha^1)\Big]\xi(y_\alpha^2)d(0,0)\nonumber\\
\bm{\Xi}^2_{\alpha\beta}\lineup = \Big[X(y_\alpha^1)-d\xi(y_\alpha^1)\Big]\xi(y_\alpha^2)d(0,0)+\Big[X(y_\beta^1)-d\xi(y_\beta^1)\Big]\xi(y_\alpha^2)d(1,0)\nonumber\\
\lineup\ \ \ +\Big[X(y_\alpha^1)-d\xi(y_\alpha^1)\Big]\xi(y_\beta^2)d(0,1)+\Big[X(y_\beta^1)-d\xi(y_\beta^1)\Big]\xi(y_\beta^2)d(1,1)\nonumber\\
\lineup\ \ \ -\Big[\xi(y_\beta^1)-\xi(y_\alpha^1)\Big]\xi(y_\alpha^2)d^1(0,0)-\Big[\xi(y_\beta^1)-\xi(y_\alpha^1)\Big]\xi(y_\beta^2)d^1(0,1)\ \ .
\end{eqnarray}
The gauge lattice measure on faces of codimension $2$ is given in \eq{0Xiabg}-\eq{1Xiabg}, after subtracting $d\xi$ from $X$. With this we can determine the operator insertions of the gauge amplitude:
\begin{eqnarray}
\Xi^2_\alpha \lineup = \ \ \ \bm{\Xi}^2_\alpha[\widetilde{C}^2_\alpha] \ \ \ \ \ = \Big[X(y_\alpha^1)-d\xi(y_\alpha^1)\Big]\xi(y_\alpha^2)\nonumber\\
\Xi^2_{\alpha\beta}\lineup = \ \ \bm{\Xi}^2_{\alpha\beta}[\widetilde{C}^2_{\alpha\beta}] \ \ \ = -\Big[\xi(y_\beta^1)-\xi(y_\alpha^1)\Big]\xi(y_\alpha^2)\nonumber\\
\Xi^2_{\alpha\beta\gamma}\lineup = \ \bm{\Xi}^2_{\alpha\beta\gamma}[\widetilde{C}^2_{\alpha\beta\gamma}] \ \,= 0\nonumber\\
\Xi^2_{\alpha\beta\gamma\delta}\lineup = \bm{\Xi}^2_{\alpha\beta\gamma\delta}[\widetilde{C}^2_{\alpha\beta\gamma\delta}] = 0\nonumber\\
\lineup\ \ \ \ \ \ \ \vdots\ \ .
\end{eqnarray}
The vertical corrections are determined by substituting into 
\begin{eqnarray}
X_\alpha^2 \lineup = (Q-d)\Xi_\alpha^2 \nonumber\\
X_{\alpha\beta}^2\lineup = (Q-d)\Xi_{\alpha\beta}^2+\Xi_\beta^2-\Xi_\alpha^2\nonumber\\
X_{\alpha\beta\gamma}^2\lineup = (Q-d)\Xi_{\alpha\beta\gamma}^2 + \Xi_{\beta\gamma}^2 - \Xi_{\alpha\gamma}^2 + \Xi_{\alpha\beta}^2\nonumber\\
X_{\alpha\beta\gamma\delta}^2\lineup = (Q-d)\Xi_{\alpha\beta\gamma\delta}^2 + \Xi_{\beta\gamma\delta}^2 - \Xi_{\alpha\gamma\delta}^2 +\Xi_{\alpha\beta\delta}^2 -\Xi_{\alpha\beta\gamma}^2\nonumber\\
\lineup\vdots\ \ .
\end{eqnarray}
This gives
\begin{eqnarray}
X^2_\alpha\lineup = \Big[X(y_\alpha^1)-d\xi(y_\alpha^1)\Big]\Big[X(y_\alpha^2)-d\xi(y_\alpha^2)\Big]\nonumber\\
X^2_{\alpha\beta} \lineup = \Big[X(y_\beta^1)-d\xi(y_\beta^1)\Big]\Big[\xi(y_\beta^2)-\xi(y_\alpha^2)\Big]+\Big[\xi(y_\beta^1)-\xi(y_\alpha^1)\Big]\Big[X(y_\alpha^2)-d\xi(y_\alpha^2)\Big]\nonumber\\
X^2_{\alpha\beta\gamma}\lineup = -\Big[\xi(y_\gamma^1)-\xi(y_\beta^1)\Big]\Big[\xi(y_\beta^2)-\xi(y_\alpha^2)\Big] \nonumber\\
X^2_{\alpha\beta\gamma\delta}\lineup = 0\nonumber\\
\lineup \vdots\ \ .\label{eq:vertX2}
\end{eqnarray}
From the form of the lattice measure, we can read off the chains defining the form of the Sen-Witten vertical corrections:
\begin{eqnarray}
C_\alpha^2 \lineup = \ell(0,0)\nonumber\\
C_{\alpha\beta}^2\lineup = \ell_1(0,0)+\ell_2(1,0)\nonumber\\
C_{\alpha\beta\gamma}^2\lineup = \ell_{12}(1,0)\nonumber\\
C_{\alpha\beta\gamma\delta}^2\lineup = 0\nonumber\\
\lineup \vdots\ \ .
\end{eqnarray}
If we drop links parallel to the $2$-axis, we recover $\widetilde{C}^2_{\alpha_0...\alpha_k}$ as expected. To give an example of computation with lattice chains, let us check that these sew together consistently with \eq{Cconst2}. From \eq{bdry} we find
\begin{eqnarray}
\d C_{\alpha\beta\gamma}^2 \lineup = \d \ell_{12}(1,0)\nonumber\\
\lineup = \ell_2(2,0)-\ell_2(1,0)-\ell_1(1,1)+\ell_1(1,0).
\end{eqnarray}
Next we need the inclusion maps of the chains of $M_{\alpha\beta},M_{\alpha\gamma}$ and $M_{\beta\gamma}$ into the lattice of $M_{\alpha\beta\gamma}$:
\begin{eqnarray}
\iota\circ C_{\alpha\beta}^2 \lineup = \ell_1(0,0)+\ell_2(1,0)\nonumber\\
\iota\circ C_{\alpha\gamma}^2 \lineup = \ell_1(0,0)+\ell_1(1,0)+\ell_2(2,0)+\ell_2(2,1)\nonumber\\
\iota\circ C_{\beta\gamma}^2 \lineup = \ell_1(1,1)+\ell_2(2,1).
\end{eqnarray}
We then find
\begin{eqnarray}
\iota\circ C_{\beta\gamma}^2 -\iota\circ C_{\alpha\gamma}^2+\iota\circ C_{\alpha\beta}^2 \lineup = \Big(\ell_1(1,1)+\ell_2(2,1)\Big)-\Big(\ell_1(0,0)+\ell_1(1,0)+\ell_2(2,0)+\ell_2(2,1)\Big)+\Big(\ell_1(0,0)+\ell_2(1,0)\Big)\nonumber\\
\lineup = -\ell_2(2,0)+\ell_2(1,0)+\ell_1(1,1)-\ell_1(1,0)
\end{eqnarray}
and
\begin{equation}
\d C_{\alpha\beta\gamma}^2=-\iota\circ C_{\beta\gamma}^2 +\iota\circ C_{\alpha\gamma}^2-\iota\circ C_{\alpha\beta}^2 ,\label{eq:Cconstex}
\end{equation}
consistent with \eq{Cconst2}. This is also shown pictorially in figure \ref{fig:ex}. 

Let us compare the expressions \eq{vertX2} to the result \eq{X2} obtained in subsection \ref{subsec:examples}. It is clear that the vertical corrections agree of we place the $\xi$ insertion on the codimension 1 faces according to
\begin{equation}y_{\alpha\beta}^2 = y_\alpha^2.\end{equation}
The codimension 2 vertical correction in \eq{X2} can be written
\begin{eqnarray}
X^2_{\alpha\beta\gamma}\lineup = \xi(y^2_{\beta\gamma})\Big[\xi(y^1_\gamma)-\xi(y^1_\beta)\Big]-\xi(y^2_{\alpha\gamma})\Big[\xi(y^1_\gamma)-\xi(y^1_\alpha)\Big]+\xi(y^2_{\alpha\beta})\Big[\xi(y^1_\beta)-\xi(y^1_\alpha)\Big]\nonumber\\
\lineup = \xi(y^2_{\beta})\Big[\xi(y^1_\gamma)-\xi(y^1_\beta)\Big]-\xi(y^2_{\alpha})\Big[\xi(y^1_\gamma)-\xi(y^1_\alpha)\Big]+\xi(y^2_{\alpha})\Big[\xi(y^1_\beta)-\xi(y^1_\alpha)\Big]\nonumber\\
\lineup = \xi(y^2_{\beta})\Big[\xi(y^1_\gamma)-\xi(y^1_\beta)\Big]-\xi(y^2_{\alpha})\Big[\xi(y^1_\gamma)-\xi(y^1_\beta)\Big]\nonumber\\
\lineup = \Big[\xi(y^2_\beta)-\xi(y^2_\alpha)\Big]\Big[\xi(y^1_\gamma)-\xi(y^1_\beta)\Big],
\end{eqnarray}
consistent with \eq{vertX2}. In this case, the vertical corrections are expressed in a fairly natural way in terms of differences of $\xi$s.

\vspace{1cm}

\noindent{\bf Acknowledgments}

\vspace{.25cm}

\noindent We would like to thank A. Sen for conversations. The work of TE is supported by ERDF and M\v{S}MT (Project CoGraDS -CZ.02.1.01/0.0/0.0/15\_ 003/0000437) .

\end{document}